\documentclass[twocolumn]{aastex631}

\usepackage{amsmath}
\newcommand{\Oiii}{[\ion{O}{3}]}
\newcommand{\Oii}{[\ion{O}{2}]}
\newcommand{\Nii}{[\ion{N}{2}]}
\newcommand{\Sii}{[\ion{S}{2}]}
\newcommand{\Ni}{[\ion{N}{1}]}
\newcommand{\Ha}{H$\alpha$}
\newcommand{\Hb}{H$\beta$}
\newcommand{\vhel}{$V_{\rm helio}$}
\newcommand{\vlsr}{$V_{\rm LSR}$}

\providecommand{\kms}{\ensuremath{\mathrm{km\,s^{-1}}}}

\graphicspath{{./}}

\shorttitle{The Silk Nebula revisited}
\shortauthors{[Author Name]}

\begin{document}

\title{The Silk Nebula Revisited: A Comment on Wu et al.\ (2026) and New Constraints from Archival Survey Spectroscopy}

\author[0000-0003-4214-8823]{Zhuo-Xiao Wang}
\affiliation{Department of Astronomy, Tsinghua University, Beijing 100084, People’s Republic of China}
\email{zhuoxiaowang@gmail.com}

\begin{abstract}
\citet{Wu2026} used one LAMOST MRS-N plate to suggest that the Silk Nebula, a $17^\circ$-long \Oiii~$\lambda$5007 filament in Gemini, is possibly associated with a supernova remnant at a fiducial distance of $\sim$160\,pc. We test this against their published values and independent archival data. The \Ni\ airglow in their blue-arm spectrum lies at air, topocentric rest wavelengths, so their \Oiii\ velocities omit the heliocentric correction and are $10.1\pm1.0$\,\kms\ too high. Corrected and combined with DESI DR1 and SDSS spectra, \Oiii\ lies at \vhel\ $=+13.3\pm1.0$\,\kms, the velocity of \Nii\ and of the local warm ionized medium (WIM). Their \Ha\ velocities are 4.5--6.7\,\kms\ redder than \Nii. Their tabulated distances are in kpc, not units of 100\,pc, and kinematic distances carry no information in this direction. Their surface brightness exceeds independent measurements (1.4--2.5 rayleighs) by factors of 4.6--8.6. Their red-line shock criteria are met equally in their nebula-free background region and describe the WIM; their other remnant diagnostics do not constrain the filament. DESI and SDSS spectra show that the filament is Balmer-dark (\Oiii/\Hb\ $>11$) and \Oii-strong, with \ion{He}{2}/\Oiii\ $<0.026$; GALEX shows a coincident ultraviolet ridge, with no \Ha, \ion{H}{1}, dust or X-ray counterpart. These properties, and the lack of an ionizing source, exclude photoionization equilibrium and complete radiative shocks and match a column-truncated radiative shock of $\sim$80--120\,\kms\ seen edge-on, plausibly on the limb of the Monogem Ring at $\sim$250--320\,pc. Far-ultraviolet and deep optical spectroscopy, a two-epoch proper motion and high-ion absorption toward bracketing white dwarfs would be decisive.
\end{abstract}

\keywords{Supernova remnants (1667) --- Emission nebulae (461) --- Interstellar medium (847) --- Spectroscopy (1558)}


\section{Introduction}\label{sec:intro}

\begin{figure*}
\centering
\includegraphics[width=0.9\textwidth]{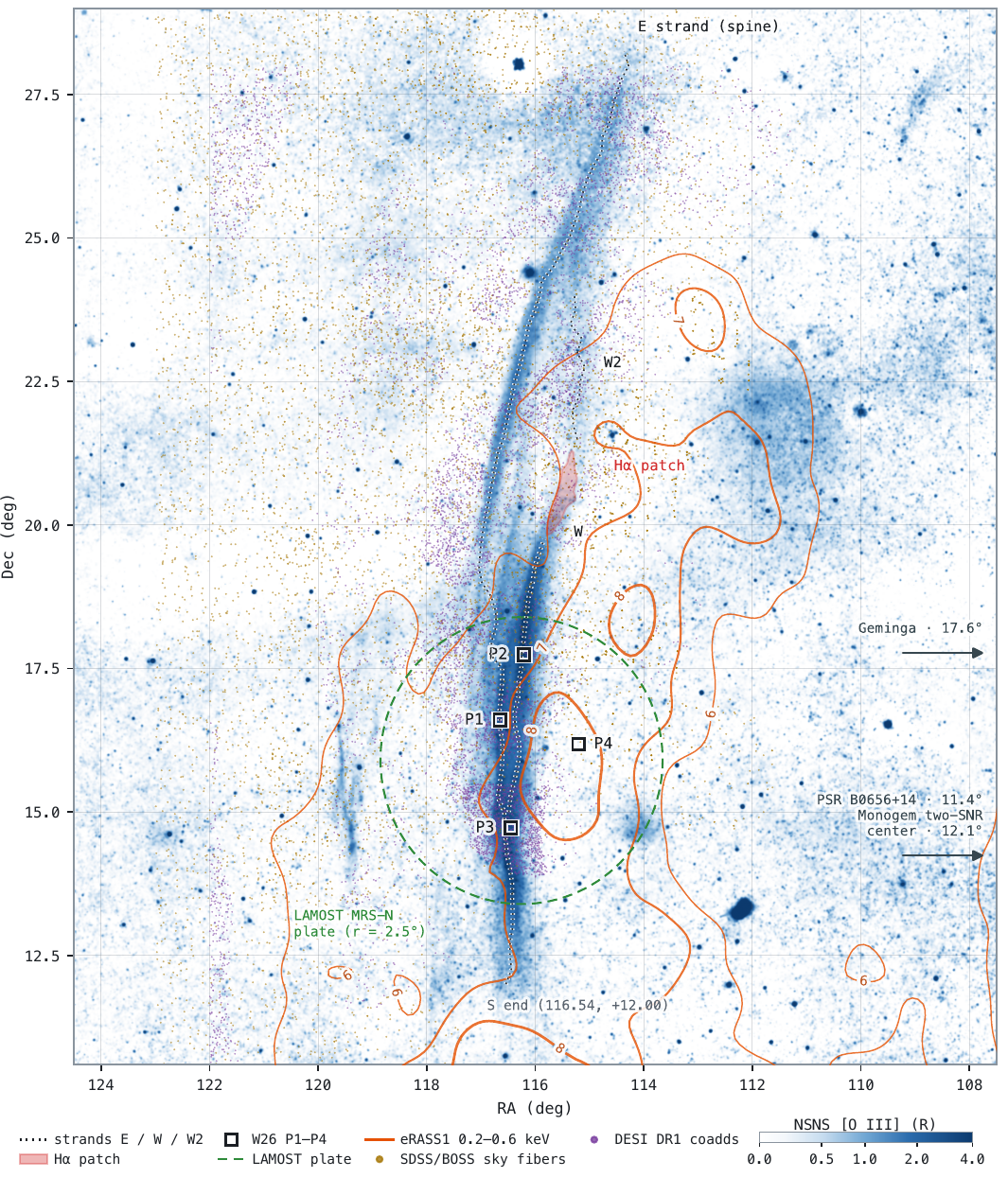}
\caption{The Silk Nebula on the NSNS DR0.2 \Oiii~$\lambda$5007 map \citep{Ziegenbalg2025}, in rayleighs (R; arcsinh scale). Dotted lines: the E (spine), W and W2 strands. Shaded polygon: half-maximum outline of the \Ha\ patch in NSNS \Ha\ imaging. Squares: regions P1--P4 of W26 \citep{Wu2026}; dashed circle: the LAMOST MRS-N plate ($2.5^\circ$ radius). Contours: 0.2--0.6\,keV surface brightness of the first SRG/eROSITA all-sky survey (eRASS1; \citealt{Merloni2024}) at 6, 7 and 8\,counts\,s$^{-1}$\,deg$^{-2}$, after masking point sources and smoothing. Dots: the SDSS/BOSS sky fibers \citep{Abdurrouf2022,SDSS2026} and DESI DR1 coadds \citep{DESI2026} analyzed here. Arrows point toward Geminga, and toward PSR B0656+14 and the two-remnant Monogem center of \citet{Knies2024}; labels give the mean angular distance of the spine from Geminga, and of its southern part (Dec $<17.6^\circ$) from the other two.}
\label{fig:overview}
\end{figure*}

The Silk Nebula is a faint \Oiii~$\lambda$5007 filament in Gemini (Figure~\ref{fig:overview}). Its ridge follows a $16.9^\circ$ arc from $(\ell,b)\approx(208^\circ,+17.5^\circ)$ to $(191^\circ,+21.8^\circ)$ with a full width at half maximum (FWHM) of $0.28^\circ$; a second strand (W) branches west from the main strand (E, the spine) near declination (Dec) $+15^\circ$.

Several groups found the filament independently. The Super Wide-Field \Oiii\ Narrowband Survey (SWONS) team detected it in late 2023 and named it the Silk Nebula \citep{QHYCCD2024}; a team member posted the first public image on 2024 January~14.\footnote{AstroBin image \url{https://www.astrobin.com/zh47r9/}.} The Northern Sky Narrowband Survey (NSNS) recorded the full arc and gave the first journal description \citep{Ziegenbalg2025}, and \citet{Fesen2026} imaged it as part of the eastern and northern limbs of the Monogem Ring.

\citeauthor{Wu2026} (\citeyear{Wu2026}; hereafter W26) used 2,684 spectra at $R\sim7500$ from one plate of the Medium-Resolution Spectroscopic Survey of Nebulae (MRS-N; \citealt{Wu2022}) of the Large Sky Area Multi-Object Fiber Spectroscopic Telescope (LAMOST; \citealt{Cui2012}), observed on 2024 February~1, and analyzed three regions on the filament (P1--P3) and a nebula-free background (P4). The blue arm (4950--5350\,\AA) covers \Oiii~$\lambda\lambda$4959, 5007 but not \Hb; the red arm (6300--6800\,\AA) covers \Ha, \Nii\ and \Sii. W26 concludes that the filament is possibly associated with a supernova remnant (SNR) at a fiducial distance of $\sim$160\,pc, where it is $\sim$28\,pc long, with energy $(2.0\pm0.7)\times10^{51}$\,erg and age $(6.8\pm1.8)\times10^{4}$\,yr, and notes that a diffuse-ionized-gas (DIG) or ionized-bubble origin cannot be ruled out.

Independent data support two W26 results: the \Oiii\ filament is real line emission, absent from P4, and the red-arm lines trace foreground and background warm ionized medium (WIM), the DIG of W26 Sect.~4.2.

The SNR interpretation, however, rests on red-line diagnostics and the \Oiii\ morphology; the size, energy and age derived for it rest on line velocities and a kinematic distance, and W26 also quotes a surface brightness. We test each of these against W26's published values and independent data, chiefly archival spectra from the Dark Energy Spectroscopic Instrument (DESI) and the Sloan Digital Sky Survey (SDSS), which also cover \Hb, \Oii\ and \ion{He}{2}, and add new constraints on the filament's excitation and location.

Section~\ref{sec:data} describes the data and methods. Sections~\ref{sec:assess} and \ref{sec:new} assess the W26 claims (Table~\ref{tab:claims}) and present new constraints; Section~\ref{sec:future} proposes decisive observations, and Section~\ref{sec:conclusions} concludes.

\begin{deluxetable*}{llc}
\tablecaption{W26 claims assessed in this work\label{tab:claims}}
\tablehead{\colhead{Claim (W26 location)} & \colhead{Our finding} & \colhead{Section}}
\startdata
\parbox[t]{0.36\textwidth}{\raggedright \Oiii\ \vhel\ $+23.8$ to $+26.8$\,\kms, 5--15\,\kms\ above the red lines; two-layer shell (Table~4; Sects.~3.2.3, 4.2)\strut} & \parbox[t]{0.48\textwidth}{\raggedright The blue-arm scale is air and topocentric, so Table~4 \Oiii\ is $10.1\pm1.0$\,\kms\ too high; \Oiii\ lies at $+13.3\pm1.0$\,\kms, $+1.1\pm1.3$\,\kms\ from \Nii\strut} & \S\ref{sec:oiii} \\
\parbox[t]{0.36\textwidth}{\raggedright \Ha\ \vhel\ $+16.3$ to $+19.7$\,\kms, used for velocities, the velocity map and distances (Table~4; Fig.~5)\strut} & \parbox[t]{0.48\textwidth}{\raggedright \Ha\ exceeds \Nii\ by $5.5\pm1.1$\,\kms\ and WHAM by $4.8\pm0.7$\,\kms; a geocoronal over-subtraction model reproduces one-half to three-quarters\strut} & \S\ref{sec:halpha} \\
\parbox[t]{0.36\textwidth}{\raggedright Fiducial distance $\sim$160\,pc from \Oiii\ kinematic distances ``in units of 100pc'' (Table~4; Sect.~3.3)\strut} & \parbox[t]{0.48\textwidth}{\raggedright Table~4 values are reproduced only as kpc ($\sim$1.6\,kpc); rotation gives $\le3.1$\,\kms\ within 300\,pc, below peculiar motions\strut} & \S\ref{sec:distance} \\
\parbox[t]{0.36\textwidth}{\raggedright \Oiii\ 23.60--23.87\,mag\,arcsec$^{-2}$, i.e., $\sim$10--13\,R (Table~3; Sect.~4.1)\strut} & \parbox[t]{0.48\textwidth}{\raggedright Exceeds DESI, SDSS and NSNS values (1.4--2.5\,R) by factors of 4.6--8.6; no identified multiplicative systematic accounts for this\strut} & \S\ref{sec:brightness} \\
\parbox[t]{0.36\textwidth}{\raggedright Shock excitation from \Sii/\Ha\ $>0.4$, \Nii/\Ha\ and diagnostic diagrams (Abstract; Sect.~4.1; Fig.~8)\strut} & \parbox[t]{0.48\textwidth}{\raggedright Ratios in nebula-free P4 equal those on the filament within $1\sigma$ and match faint high-latitude WIM\strut} & \S\ref{sec:shock} \\
\parbox[t]{0.36\textwidth}{\raggedright $n_e$, $T_e$, $V_t$, O/H, \Oiii\ widths, energetics and radio non-detection (Sects.~3.2--4.1; Tables~2, 4)\strut} & \parbox[t]{0.48\textwidth}{\raggedright Table~2 ratios imply $n_e=225$--631\,cm$^{-3}$; $T_e$ ($\sigma_T=7300$--10700\,K), $V_t$ and O/H are consistent with P4; widths not recoverable at S/N~6; energetics inputs unsupported; NVSS/TGSS resolve the filament out\strut} & \S\ref{sec:snrdiag}
\enddata
\tablecomments{Table, figure and section numbers in the first two columns refer to W26. Velocities are heliocentric. P4 is the nebula-free background region of W26. WHAM, Wisconsin H$\alpha$ Mapper; WIM, warm ionized medium; $n_e$, electron density; $T_e$, electron temperature ($\sigma_T$, its $1\sigma$ uncertainty); $V_t$, turbulent velocity.}
\end{deluxetable*}

\section{Data and methods}\label{sec:data}

\subsection{W26 material}

We use W26 arXiv version~1, whose tables match the journal version. We extracted the W26 Fig.~3 spectra and the \Ha\ \vlsr\ map of W26 Fig.~5 from the PDF vector data (spectral axes calibrated to 0.002\,\AA\ rms), and digitized the \Oiii\ \vlsr\ histogram of the same figure. The per-fiber MRS-N nebular spectra are not public \citep{Zhang2025}, so our corrections to W26 values apply in aggregate.

\subsection{Archival spectra}

\emph{DESI DR1.} The DESI pipeline \citep{Guy2023} subtracts a sky model built from the sky fibers of each petal and exposure, so the filament leaves positive residuals on the ridge and negative ones elsewhere on its petals. In $\pm40$\,\AA\ line windows of 7,620 DR1 coadds \citep{DESI2026} near the filament and in control fields $>5^\circ$ away, we subtracted each target's pipeline model and regressed each line residual on that predicted for Gaussian ridges along the traced strands, minus the tile--petal mean (a proxy for the sky-fiber mean). This isolates emission correlated with the filament, canceling smooth WIM and sky, and gives the \Oiii\ intensity and line ratios relative to \Oiii. HEALPix block-bootstrap uncertainties are inflated to match line-free null windows, and the [\ion{O}{1}]~$\lambda$5577 sky line gauges template-correlated systematics. Surface brightness uses the per-exposure guide-camera point-source fiber acceptance (median 0.566 at 5008\,\AA; $\pm8\%$ systematic, or $\approx$20\% for two survey-validation tiles that lack guide-camera data; Appendix~\ref{sec:appendix-desi}).

\emph{SDSS sky fibers.} We used 9,092 sky-fiber spectra from SDSS-I/II, BOSS and eBOSS (DR17; \citealt{Abdurrouf2022}) and SDSS-V (DR19; \citealt{SDSS2026}), in 302 plate--night--spectrograph groups (250 dark). Because \Oiii~$\lambda$5007 and \Hb\ have no significant airglow, the total sky (pipeline residual plus sky model) measures them absolutely. ON$-$OFF differences within each group (ON within $0.15^\circ$ of a strand, OFF at $0.45^\circ$--$2.0^\circ$) cancel plate-level systematics, [\ion{O}{1}]~$\lambda$5577 sets the velocity zero points, and shifted-trace placebos calibrate residual systematics. Matching to SDSS imaging sky levels \citep{Almeida2023} gives diffuse-light correction factors of $0.84\pm0.08$ for 3\arcsec\ and $0.60\pm0.06$ for 2\arcsec\ fibers.

\subsection{Imaging and ancillary data}

We traced the ridge (66 points at $0.25^\circ$ declination steps; median uncertainty $0.022^\circ$) on W26 Figs.~1 and 4 and the NSNS DR0.2 maps \citep{Ziegenbalg2025}, registered to Tycho-2 stars \citep{Hog2000}; the NSNS calibration is adopted as published. Ancillary data are GALEX far- and near-ultraviolet (FUV, NUV) images \citep{Martin2005,Morrissey2007}, in continuum units (CU; photons\,cm$^{-2}$\,s$^{-1}$\,sr$^{-1}$\,\AA$^{-1}$); \Ha\ from the Southern H-Alpha Sky Survey Atlas (SHASSA; \citealt{Gaustad2001}) and the Wisconsin \Ha\ Mapper (WHAM; \citealt{Haffner2003}); the Galactic Arecibo L-band Feed Array \ion{H}{1} survey (GALFA-\ion{H}{1}) DR2 \citep{Peek2018}; the IRIS reprocessing of the IRAS 100\,$\mu$m maps \citep{MivilleDeschenes2005}; SRG/eROSITA eRASS1 0.2--0.6\,keV maps \citep{Merloni2024}; and the FIMS/SPEAR \ion{C}{4} map of \citet{Kim2007}, digitized from the published figure. Counterpart tests use cross-filament profiles and template regressions on the \Oiii\ ridge, calibrated with shifted and rotated traces.

\subsection{Kinematics and conventions}

To test the W26 kinematic distances we re-implemented the \texttt{kd} package \citep{Wenger2018}: its \citet{Persic1996} rotation curve with \citet{Reid2014} parameters, and its Monte Carlo estimator.

Velocities are heliocentric (\vhel; equal to barycentric within 0.02\,\kms) unless given as \vlsr, for which we use, as W26, the standard solar motion of 20\,\kms\ toward $\alpha=18^{\rm h}03^{\rm m}50\fs29$, $\delta=+30^\circ00'16\farcs8$ (J2000). Then \vlsr$\,-\,$\vhel\ is $-12.10$\,\kms\ at P1 and ranges from $-13.2$ to $-9.3$\,\kms\ along the spine; at P1--P4 the solar motion of \citet{Schonrich2010} changes it by $\le0.4$\,\kms. We checked the velocity frame of each spectroscopic dataset against sky lines or header corrections. Wavelengths are vacuum; line names use air wavelengths. Surface brightness is in rayleighs, $1\,{\rm R}=10^6/(4\pi)$\,photons\,cm$^{-2}$\,s$^{-1}$\,sr$^{-1}$, or $7.42\times10^{-18}$\,erg\,cm$^{-2}$\,s$^{-1}$\,arcsec$^{-2}$ at \Oiii~$\lambda$5007. Line ratios are energy ratios unless marked photon. Uncertainties are $1\sigma$ (or 68\% intervals) and limits $2\sigma$ unless stated.

Derived data and analysis code are available at [repository URL].

\section{Assessment of the W26 analysis}\label{sec:assess}

Table~\ref{tab:claims} lists the W26 claims that we re-examine and our findings. We begin with the velocities, on which the distance and the two-layer shell model depend; Table~\ref{tab:velocity} lists them by region and dataset.

\subsection{The \texorpdfstring{\Oiii{}}{[O III]} velocity offset is a wavelength-reference artifact}\label{sec:oiii}

W26 states that the LAMOST pipeline corrected the MRS-N spectra for heliocentric motion (W26 Sect.~3.2.3) and quotes a radial-velocity precision of $<1.5$\,\kms{} (W26 Sect.~2.2). W26 Table~4 places \Oiii{} at $+23.8$ to $+26.8$\,\kms{} (P1--P3) but \Nii{} and \Sii{} at $+11.7$ to $+13.3$\,\kms{} (P1--P4). This offset of $\sim$5--15\,\kms{} underlies the 100--150\,pc discrepancy between blue- and red-arm distances, the two-layer shell interpretation (W26 Sect.~4.2) and the \Oiii-only kinematic distance (W26 Sect.~3.3).

\begin{figure*}
\includegraphics[width=\textwidth]{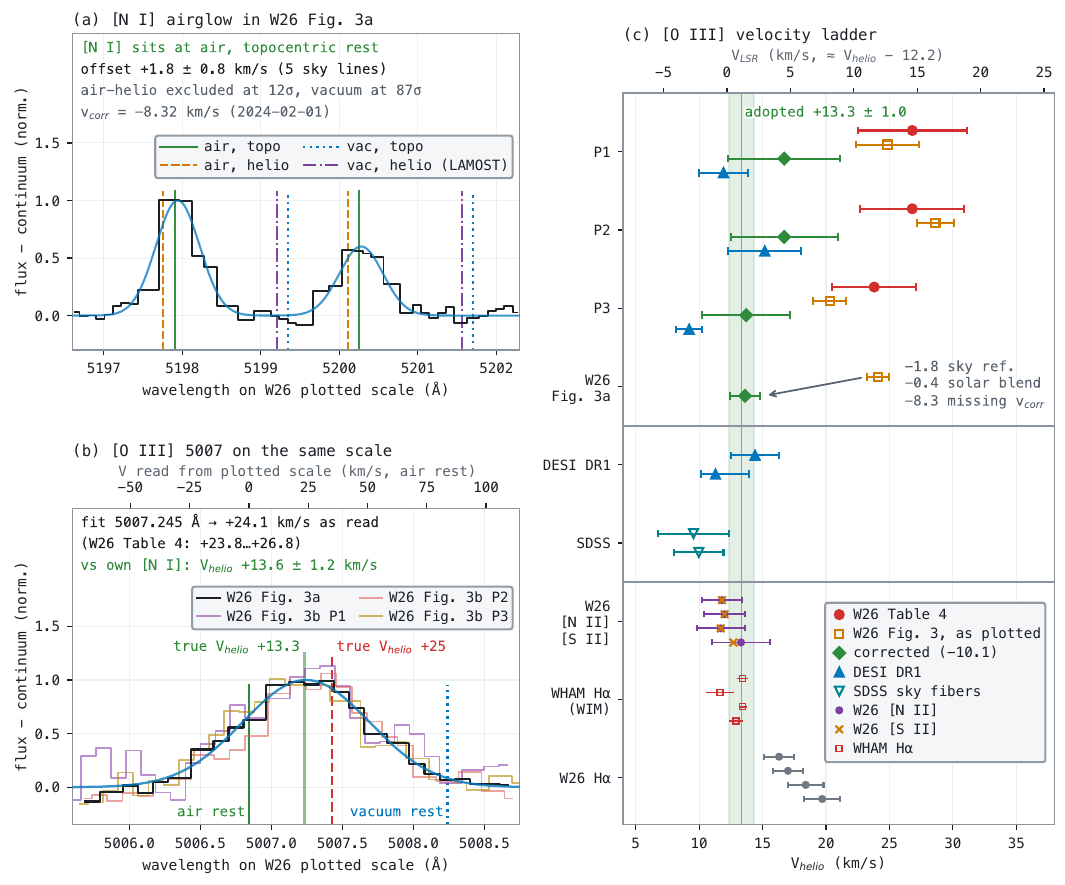}
\caption{Wavelength frame of the blue-arm spectra re-extracted from W26 \citep{Wu2026} Fig.~3, and the resulting \Oiii{} velocities. (a) \Ni~$\lambda\lambda$5198, 5200 airglow in W26 Fig.~3a with a two-Gaussian fit. Lines: doublet positions on air or vacuum, topocentric or heliocentric scales. (b) \Oiii~$\lambda$5007 in W26 Fig.~3a (with Gaussian fit) and W26 Fig.~3b (P1--P3), normalized. Lines: air and vacuum rest wavelengths, and plotted positions for true \vhel{} $=+13.3$ and $+25$\,\kms; top axis: plotted-scale velocity. (c) \vhel{} (top axis: \vlsr{} $\approx$ \vhel{} $-\,12.2$\,\kms). P1--P3: W26 Table~4, W26 Fig.~3b as plotted, W26 Table~4 minus 10.1\,\kms, and DESI DR1 within $0.5\arcdeg$. Other rows: W26 Fig.~3a as plotted and referenced to its airglow; DESI \citep{DESI2026} and SDSS \citep{Abdurrouf2022} estimates (Table~\ref{tab:velocity}, note a); and, for P1--P4, W26 \Nii, \Sii{} and \Ha{} and nearest-beam WHAM \Ha{} (\citealt{Haffner2003}; Table~\ref{tab:velocity}). Band: adopted value.\label{fig:oiii}}
\end{figure*}

The W26 Fig.~3a blue-arm spectrum (Section~\ref{sec:data}) contains the \Ni~$\lambda\lambda$5198, 5200 airglow doublet. On the plotted scale, five sky lines (the doublet and three weaker lines) lie $+1.8\pm0.8$\,\kms{} from their air rest wavelengths \citep{Hanuschik2003,Kramida2024}. An air, heliocentric scale would place them at $-8.3$\,\kms{} (excluded at 12$\sigma$) and a vacuum scale at $+75$ to $+83$\,\kms{} ($\geq87\sigma$; Figure~\ref{fig:oiii}a). The plotted scale is therefore air and topocentric, and lacks the heliocentric correction ($v_{\rm corr}=-8.32$\,\kms{} for the 2024 February~1 exposures; \citealt{Astropy2022}), although the LAMOST pipeline delivers vacuum, heliocentric wavelengths \citep{Luo2015,Ren2021}. W26 Table~4 was measured on the same scale: the W26 Fig.~3b spectra reproduce it to $-1.5\pm2.4$\,\kms, whereas a heliocentric table would lie 10.1\,\kms{} below them (excluded at 4.5$\sigma$).

Measured against the airglow in the same spectrum, \Oiii{} needs no wavelength calibration. In W26 Fig.~3a, \Oiii~$\lambda$5007 lies at $+24.1\pm0.9$\,\kms{} on the plotted scale; removing the $+1.8$\,\kms{} sky offset and a $+0.4$\,\kms{} bias from a solar Fraunhofer blend \citep{Kurucz2005}, and adding $v_{\rm corr}$, gives \vhel{} $=+13.6\pm1.2$\,\kms. The W26 Table~4 \Oiii{} velocities are therefore too high by $10.1\pm1.0$\,\kms, the omitted $v_{\rm corr}$ plus the sky offset. Only W26 Fig.~3a shows the airglow, so the correction is applied in aggregate.

Independent spectra agree (Figure~\ref{fig:oiii}c). DESI DR1 \citep{DESI2026}, whose spectra are vacuum and barycentric \citep{Guy2023}, gives $+14.4\pm1.9$\,\kms{} from 390 on-ridge fibers, and SDSS sky fibers \citep{Abdurrouf2022} give $+9.6\pm2.8$\,\kms. DESI and SDSS alone give $+12.9\pm1.6$\,\kms; with the airglow-referenced value, the weighted mean is \vhel{} $=+13.3\pm1.0$\,\kms{} (\vlsr{} $=+1.2\pm1.0$\,\kms; $\chi^2=2.2$ for 2 degrees of freedom), with a further $\pm1$\,\kms{} estimator systematic. \Oiii{} $-$ \Nii{} becomes $+1.1\pm1.3$\,\kms, against $+13.7\pm2.4$\,\kms{} for W26 P1--P3, and the WIM \Ha{} observed by WHAM \citep{Haffner2003} on the southern filament lies at $+13.8\pm0.8$\,\kms.

The LAMOST pipeline calibration is not the cause. On the W26 plate, LAMOST DR12 \citep{Cui2012} stellar velocities give blue minus red arm $=+0.7\pm0.5$\,\kms{} (76 stars), and both arms agree with Gaia DR3 for six stars \citep{Gaia2023,Katz2023}. The simplest explanation is that the blue arm was re-referenced to air sky-line wavelengths after the pipeline, which removed the heliocentric correction; \citet{Ren2021} and \citet{Wu2022} document sky-line recalibration of the red arm only. The correction does not depend on this mechanism.

The \Oiii--red-line offset is therefore not a property of the nebula, and the quoted $<1.5$\,\kms{} precision does not apply to the published blue-arm velocities. The two-layer shell and inhomogeneous-bubble scenarios and the blue/red distance discrepancy are not supported, and the \Oiii{} kinematic distances rest on \vlsr{} $=+11$ to $+16$\,\kms{} instead of the corrected $+1$ to $+5$\,\kms{} (Section~\ref{sec:distance}). Velocity does not separate the \Oiii{} filament from the WIM.

\subsection{The \texorpdfstring{\Ha{}}{H-alpha} velocities are red-biased}\label{sec:halpha}

\begin{figure*}
\includegraphics[width=\textwidth]{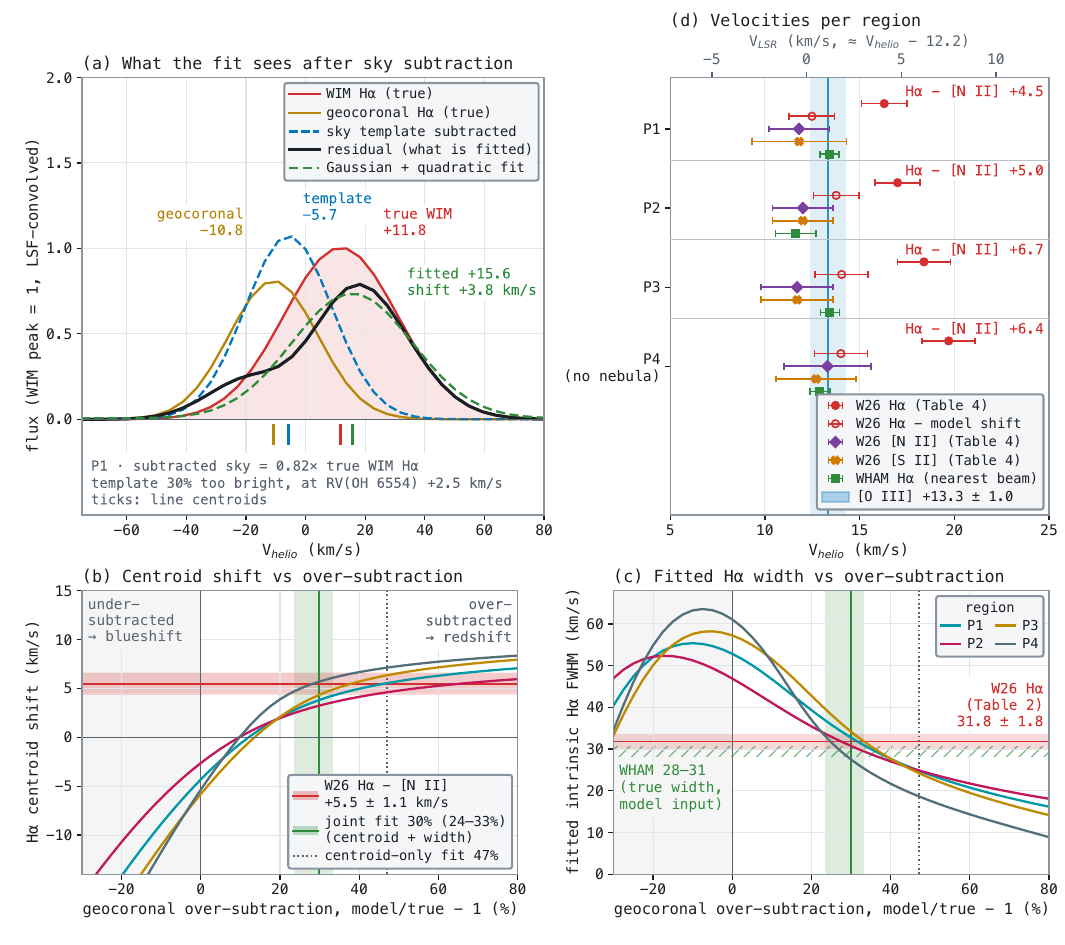}
\caption{Forward model of geocoronal \Ha{} over-subtraction by the MRS-N sky template \citep{Zhang2021}, fitted to published W26 values \citep{Wu2026} and not to spectra, and red-line velocities by region. Profiles in (a) are convolved with the instrumental line-spread function (LSF). (a) P1 at the joint centroid-and-width fit: WIM \Ha{} (W26 \Nii{} velocity; WHAM width, \citealt{Haffner2003}), geocoronal \Ha, subtracted template (30\% too bright, $+2.5$\,\kms{} from the OH $\lambda$6554 velocity), residual and its Gaussian-plus-quadratic fit; ticks: centroids. (b) Fitted centroid shift versus over-subtraction for P1--P4 at the joint-fit template position. Horizontal bands: W26 \Ha{} $-$ \Nii{} (weighted mean; P1--P4 range); vertical band: joint fit; dotted line: centroid-only fit. (c) As (b) for the fitted intrinsic \Ha{} FWHM; band: W26 Table~2 mean; hatched: WHAM widths. (d) \vhel{} of \Ha, \Nii{} and \Sii{} (W26 Table~4), \Ha{} minus the model shift, and nearest-beam WHAM \Ha; labels: \Ha{} $-$ \Nii; band: adopted \Oiii{} velocity; top axis: \vlsr{} $\approx$ \vhel{} $-\,12.2$\,\kms.\label{fig:halpha}}
\end{figure*}

W26 Table~4 gives \Ha{} velocities of $+16.3$ to $+19.7$\,\kms{} for P1--P4, W26 Fig.~5 maps the \Ha{} \vlsr, and W26 Sect.~3.2.3 prefers \Ha{} in P4 for its smaller uncertainty. In every region \Ha{} is redder than \Nii{} and \Sii{} from the same red-arm spectra: weighted over P1--P4, \Ha{} $-$ \Nii{} $=+5.5\pm1.1$\,\kms{} (4.9$\sigma$), whereas \Nii{} $-$ \Sii{} $=+0.1\pm1.4$\,\kms{} (Figure~\ref{fig:halpha}d). Relative to the nearest WHAM beams, \Ha{} $-$ WHAM $=+4.8\pm0.7$\,\kms{} but \Nii{} $-$ WHAM $=-0.8\pm1.0$\,\kms. The offset is equally large in the nebula-free P4.

A frame error would shift \Nii{} and \Sii{} as well, and a wavelength-scale error would be largest for \Sii, the line farthest from the red-arm sky-line calibration range of 6287--6553\,\AA{} (air; \citealt{Ren2021,Wu2022}); yet \Sii{} agrees with \Nii{} and WHAM. An erroneous \Ha{} rest wavelength cannot be excluded, but it would have to be $6564.49\pm0.02$\,\AA, below every \ion{H}{1} $n=2$--3 component listed by \citet{Kramida2024}, and would also be an artifact.

Geocoronal \Ha{} offers a plausible mechanism. On 2024 February~1 it lay 22--25\,\kms{} from the WIM line, within one instrumental FWHM (34.5\,\kms; \citealt{Zhang2021}). The MRS-N pipeline subtracts a sky template whose intensity is predicted from the solar altitude \citep{Zhang2021,Wu2022}; on this plate the subtracted sky is 0.86--1.27 times the published nebular \Ha. In our forward model, only over-subtraction shifts the fitted centroid redward by more than 0.4\,\kms{} (Figure~\ref{fig:halpha}a,b). A joint fit to the W26 \Ha{} offsets and widths, with the WHAM width as the true width, requires 30\% (24--33\%) over-subtraction (Figure~\ref{fig:halpha}b,c) and reproduces about three-quarters of the offset, or about one-half with the template fixed at the pipeline position. Fitting the offsets alone requires 47\% but predicts intrinsic widths of 13--23\,\kms, against 31.0--32.4\,\kms{} in W26 Table~2. These are model results, not fits to spectra: the spectra and that night's geocoronal intensity are not available.

The bias itself does not depend on the model. The \Ha{} velocities of W26 Table~4 and W26 Fig.~5 carry a systematic error of $+4.5$ to $+6.7$\,\kms{} that the quoted 1.2--1.4\,\kms{} uncertainties do not include, and the \Ha-based kinematic distances inherit it. These values need confirmation, for example by a re-reduction with the geocoronal intensity left free or the \Ha{} centroid tied to \Nii. The W26 inference of a DIG contribution to the red-arm lines (W26 Sect.~4.2) is unaffected, because the bias is the same on and off the nebula.

\begin{deluxetable*}{lccccc}
\tablecaption{Heliocentric velocities by region and dataset (km\,s$^{-1}$)\label{tab:velocity}}
\tablehead{
\colhead{Dataset, line} & \colhead{P1} & \colhead{P2} & \colhead{P3} & \colhead{P4 (no nebula)} & \colhead{Aggregate\tablenotemark{a}}
}
\startdata
W26 Table~4, \Oiii & $+26.8\pm4.3$ & $+26.8\pm4.1$ & $+23.8\pm3.3$ & \nodata & $+25.5\pm2.2$ \\
\Oiii{} corrected ($-10.1\pm1.0$) & $+16.7\pm4.4$ & $+16.7\pm4.2$ & $+13.7\pm3.5$ & \nodata & $+15.4\pm2.3$ \\
\Oiii{} vs.\ W26 \Ni{} airglow (W26 Fig.~3a) & \nodata & \nodata & \nodata & \nodata & $+13.6\pm1.2$ \\
DESI DR1, \Oiii\tablenotemark{b} & $+11.9\pm1.9$ (61) & $+15.2\pm2.9$ (9) & $+9.2\pm1.0$ (128) & \nodata & $+14.4\pm1.9$ \\
SDSS sky fibers, \Oiii & \nodata & \nodata & \nodata & \nodata & $+9.6\pm2.8$ \\
Adopted \Oiii & \nodata & \nodata & \nodata & \nodata & $+13.3\pm1.0$ \\
W26 Table~4, \Nii & $+11.8\pm1.6$ & $+12.0\pm1.6$ & $+11.7\pm1.9$ & $+13.3\pm2.3$ & $+12.2\pm0.9$ \\
W26 Table~4, \Sii & $+11.8\pm2.5$ & $+12.0\pm1.6$ & $+11.7\pm1.9$ & $+12.7\pm2.1$ & \nodata \\
WHAM, \Ha\tablenotemark{c} & $+13.4\pm0.4$ & $+11.6\pm1.1$ & $+13.4\pm0.3$ & $+12.9\pm0.5$ & $+13.8\pm0.8$ \\
W26 Table~4, \Ha & $+16.3\pm1.2$ & $+17.0\pm1.2$ & $+18.4\pm1.4$ & $+19.7\pm1.4$ & \nodata \\
\enddata
\tablecomments{W26 values are as published. \vlsr{} $=$ \vhel{} $+\,\Delta V$, with $\Delta V=-12.10$, $-11.87$, $-12.58$ and $-12.38$\,\kms{} for P1--P4 and $-12.2$\,\kms{} (mean southern value) for the aggregate column (Section~\ref{sec:data}).}
\tablenotetext{a}{W26 and corrected \Oiii: P1--P3 weighted mean. \Ni{} airglow: one spectrum of unstated position. DESI: on-ridge stack of 390 fibers, including a $\pm1.5$\,\kms{} zero-point uncertainty; a joint fit to both \Oiii{} lines gives $+11.3$ (16--84\%: 10.2--13.9). SDSS: ON$-$OFF stack of all strands; an independent reduction gives $+12.3\pm2.5$ (all strands) and $+10.0\pm1.8\pm0.8$ (Dec $<20\arcdeg$). Adopted: weighted mean of the airglow-referenced, DESI and SDSS values. \Nii: unweighted mean of P1--P4. WHAM: mean and beam-to-beam scatter of six beams on the southern filament, converted beam by beam (\vlsr{} $=+1.3$\,\kms{} as observed; mean $\Delta V=-12.5$\,\kms).}
\tablenotetext{b}{Weighted mean of positive-residual fibers with S/N $>4$ within $0.5\arcdeg$ (number in parentheses), without the zero-point uncertainty (Appendix~\ref{sec:appendix-desi}). P2 (Dec $+17.7\arcdeg$) lies at the southern edge of the zone (Dec 17.6--21$\arcdeg$) where DESI, but not SDSS, gives $+17$ to $+23$\,\kms{} (Section~\ref{sec:spectrum}).}
\tablenotetext{c}{Main component of the nearest $1\arcdeg$ beam ($0.27$--$0.48\arcdeg$ away); the uncertainty is the beam-to-beam scatter within $1\arcdeg$.}
\end{deluxetable*}

\subsection{Distance: W26 Table~4 units and the absence of kinematic distance information}
\label{sec:distance}

W26 Sect.~3.3 derives Monte Carlo and traditional kinematic distances with the \texttt{kd} code \citep{Wenger2018} and the \citet{Reid2014} rotation curve. The footnotes to W26 Table~4 give them ``in units of 100pc'', and W26 adopts ``a fiducial distance of $\sim$160\,pc'', ``the approximate median of the six \Oiii-based kinematic distance estimates''. We tested the unit using only W26's printed values and the rotation curve it names (Section~\ref{sec:data}).

\begin{figure*}
\centering
\includegraphics[width=\textwidth]{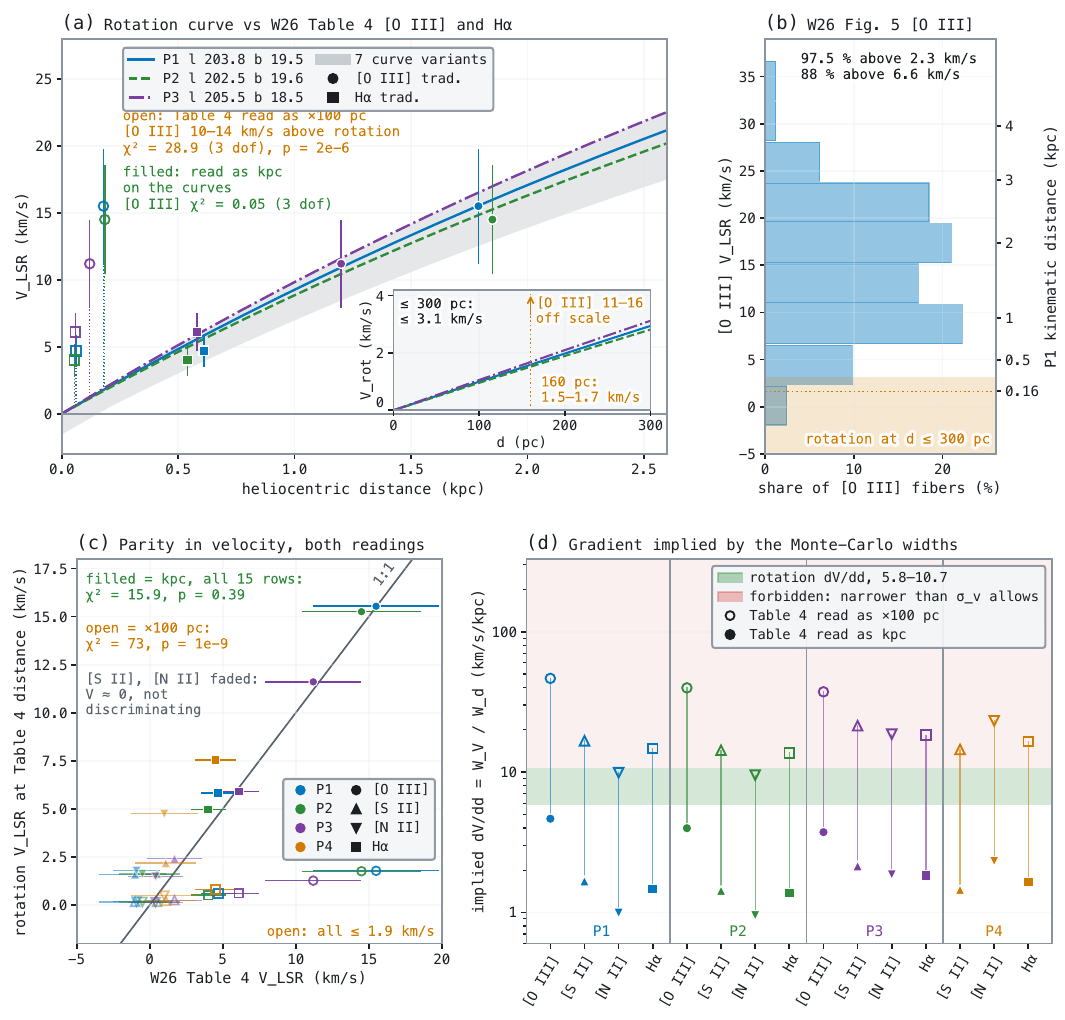}
\caption{W26 \citep{Wu2026} Table~4 distances read as kpc (filled symbols) and as 100\,pc (open symbols), compared with the \citet{Reid2014} rotation curve in its \texttt{kd} form \citep{Wenger2018}. (a) Rotation \vlsr\ versus distance toward P1--P3 (gray band, seven curve variants), with \Oiii\ and \Ha\ \vlsr\ $\pm\sigma_v$ at the traditional distances; inset, 0--300\,pc. (b) \Oiii\ \vlsr\ histogram digitized from W26 Fig.~5; shaded band, rotation within 300\,pc; dotted line, rotation at 160\,pc; right axis, P1 kinematic distance. (c) Rotation \vlsr\ at the printed distance versus printed \vlsr\ $\pm\sigma_v$ for all 15 rows (\Sii\ and \Nii\ faded). (d) Gradient $W_V/W_d$ implied by each Monte Carlo interval ($W_d$, printed width; $W_V$, 68.3\% velocity width for the tabulated $\sigma_v$); lower band, rotation gradient (curve variants and the Oort term of \citealt{Bovy2017}); upper shading, intervals narrower than $\sigma_v$ allows.}
\label{fig:distance}
\end{figure*}

Inverting the printed \vlsr\ along the W26 Table~1 sightlines reproduces the \Oiii\ traditional distances read as kpc: we obtain 1.785 against the printed 1.79 (P1), 1.743 (1.803 from the printed \vhel) against 1.85 (P2), and 1.154 against 1.20 (P3). Comparing each printed \vlsr\ with the rotation velocity at the printed distance gives $\chi^2=0.05$ for the three \Oiii\ rows in kpc and $\chi^2=28.9$ (3 degrees of freedom, $p=2\times10^{-6}$) in units of 100\,pc (Figure~\ref{fig:distance}a); for all 15 rows the values are 15.9 ($p=0.39$) and 73.0 ($p=1\times10^{-9}$; Figure~\ref{fig:distance}c). Seven curve variants, including that of \citet{Reid2019}, give \Oiii\ distances at least 9 times the 100-pc reading. At 160\,pc, rotation produces only 1.5--1.7\,\kms\ toward P1--P3, whereas 97.5\% of the \Oiii\ velocities in the W26 Fig.~5 histogram exceed $+2.3$\,\kms, which requires $d>0.23$\,kpc on the P1 curve (Figure~\ref{fig:distance}a,b).

Two further properties of W26 Table~4 require kpc. First, a velocity width $W_V$ maps to a Monte Carlo interval of width $W_d\approx W_V/({\rm d}V/{\rm d}d)$, which other error terms only widen, so for the tabulated velocity errors $W_V/W_d$ cannot exceed the rotation gradient, 5.8--10.7\,\kms\,kpc$^{-1}$ along P1--P4. Read as 100\,pc, the \Oiii\ intervals imply 37--47\,\kms\,kpc$^{-1}$, and 13 of 15 rows exceed the bound; read as kpc, all 15 lie below it (Figure~\ref{fig:distance}d). Second, the Monte Carlo lower bounds of all 12 red-line rows, 0.00--0.02, match the 0.01-kpc floor of the \texttt{kd} distance grid; read as 100\,pc they would lie below it.

The fiducial 160\,pc is 100 times the median, 1.605, of the six \Oiii\ values (1.63, 1.58, 0.98, 1.79, 1.85, 1.20). The kinematic distance implied by W26's own analysis is therefore $\sim$1.6\,kpc, and the $\sim$28\,pc extent (281\,pc at 1.605\,kpc), the $\sim$30\,pc radius, and the energy and age derived at 160\,pc (W26 Sects.~3.3 and 4.1) do not follow from the data.

\begin{figure*}
\centering
\includegraphics[width=\textwidth]{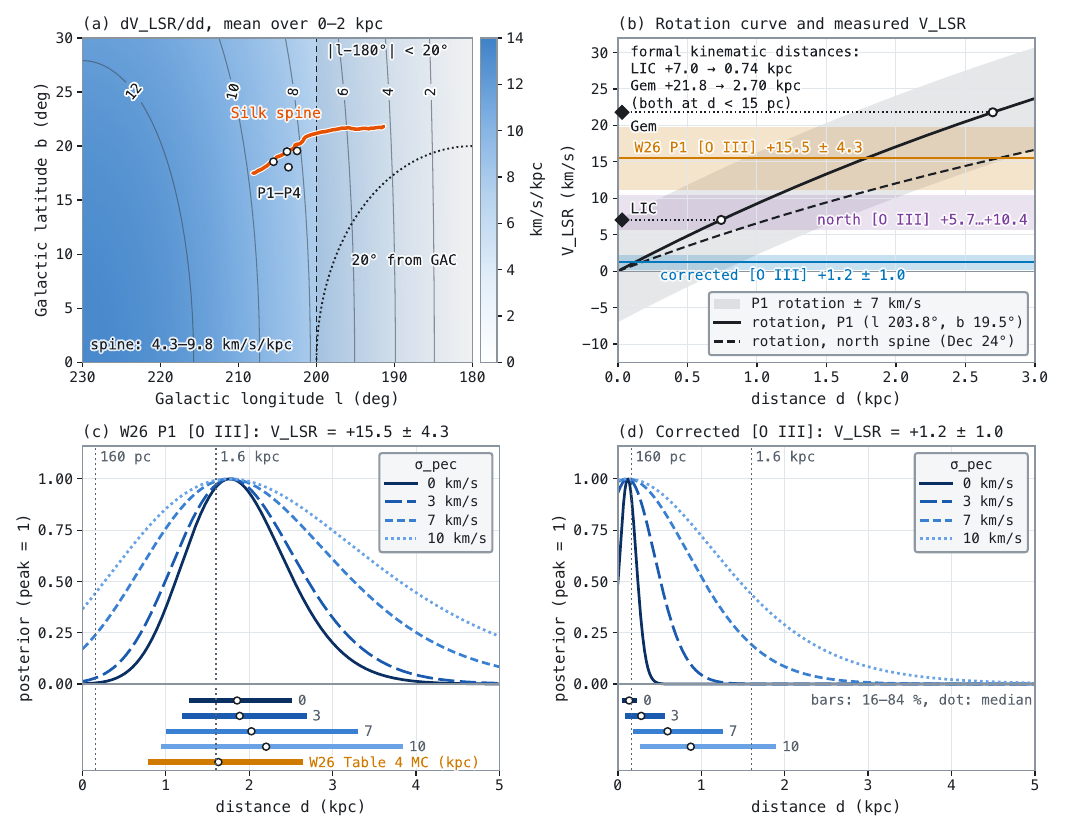}
\caption{Kinematic distances toward the Silk Nebula for the rotation curve of Figure~\ref{fig:distance}. (a) Mean rotation gradient ${\rm d}V_{\rm LSR}/{\rm d}d$ over 0--2\,kpc (contours; \kms\,kpc$^{-1}$), with the \Oiii\ spine and P1--P4 (dots); dashed line, edge of the $160^\circ<\ell<200^\circ$ band masked by \citet{Wenger2018}; dotted curve, $20^\circ$ from the Galactic anticenter (GAC). (b) Rotation \vlsr\ toward P1 (solid; gray band, $\pm7$\,\kms) and toward the spine at Dec $+24^\circ$ (dashed); horizontal bands, \Oiii\ \vlsr\ of W26 \citep{Wu2026} for P1, the corrected value (Section~\ref{sec:oiii}) and the northern DESI range (Section~\ref{sec:spectrum}); diamonds, the LIC and Gem clouds \citep{Redfield2008}; open circles, their formal kinematic distances. (c, d) Peak-normalized posteriors $P(d\,|\,V_{\rm LSR})$ toward P1 for \vlsr\ $=+15.5\pm4.3$ and $+1.2\pm1.0$\,\kms\ and $\sigma_{\rm pec}=0$, 3, 7 and 10\,\kms\ (flat prior on 0--6\,kpc; curve parameters marginalized); bars, 16--84\% intervals with medians; lowest bar in (c), W26 Table~4 Monte Carlo interval read as kpc.}
\label{fig:kinematic}
\end{figure*}

W26 Sect.~3.3 notes that the nebula lies near the boundary of the anticenter region excluded from kinematic analysis; 8.1$^\circ$ of the 16.9$^\circ$ spine lies within the $160^\circ<\ell<200^\circ$ band masked by \citet{Wenger2018} (Figure~\ref{fig:kinematic}a). In this direction the rotational signal is comparable to typical peculiar velocities: toward P1--P4 the mean rotation gradient over 0--2\,kpc is 7.6--9.1\,\kms\,kpc$^{-1}$, so peculiar velocities of 5--10\,\kms\ \citep{Stark1989,Anderson2012,Reid2019} shift the distance by 0.55--1.32\,kpc. Within 300\,pc, rotation toward P1--P4 contributes at most 3.1\,\kms, so the \vlsr\ of a nearby object is essentially its peculiar velocity. In the \citet{Redfield2008} model, the Local Interstellar Cloud (LIC) and the Gem cloud, both within 15\,pc, project to \vlsr\ $=+7.0$ and $+21.8$\,\kms\ toward P1 and would receive kinematic distances of 0.74 and 2.70\,kpc (Figure~\ref{fig:kinematic}b).

The \texttt{kd} Monte Carlo behind W26 Table~4 contains no peculiar-velocity term \citep{Wenger2018}. With a peculiar-velocity dispersion $\sigma_{\rm pec}=10$\,\kms, the 16--84\% posterior interval for the W26 P1 \Oiii\ velocity widens from 1.28--2.51 to 0.94--3.85\,kpc, and that for the corrected \vlsr\ $=+1.2\pm1.0$\,\kms\ (Section~\ref{sec:oiii}) from 0.06--0.23 to 0.26--1.90\,kpc (Figure~\ref{fig:kinematic}c,d). The posterior is thus set by the adopted $\sigma_{\rm pec}$, not by the measurement, and the W26 data do not constrain the distance; Section~\ref{sec:location} discusses non-kinematic constraints.

\subsection{Surface brightness}
\label{sec:brightness}

W26 Table~3 lists \Oiii-band surface brightnesses of 23.60, 23.87 and 23.85\,mag\,arcsec$^{-2}$ for P1--P3, which W26 Sect.~4.1 converts to ``$\sim$10--13 Rayleigh'', ``well above'' the 0.5--5\,R detection limit of the \citet{Fesen2024} images. The conversion is correct: an AB magnitude \citep{Oke1983} in a 60-\AA\ band containing only \Oiii~$\lambda$5007 gives 12.75, 9.94 and 10.12\,R.

\begin{figure*}
\centering
\includegraphics[width=\textwidth]{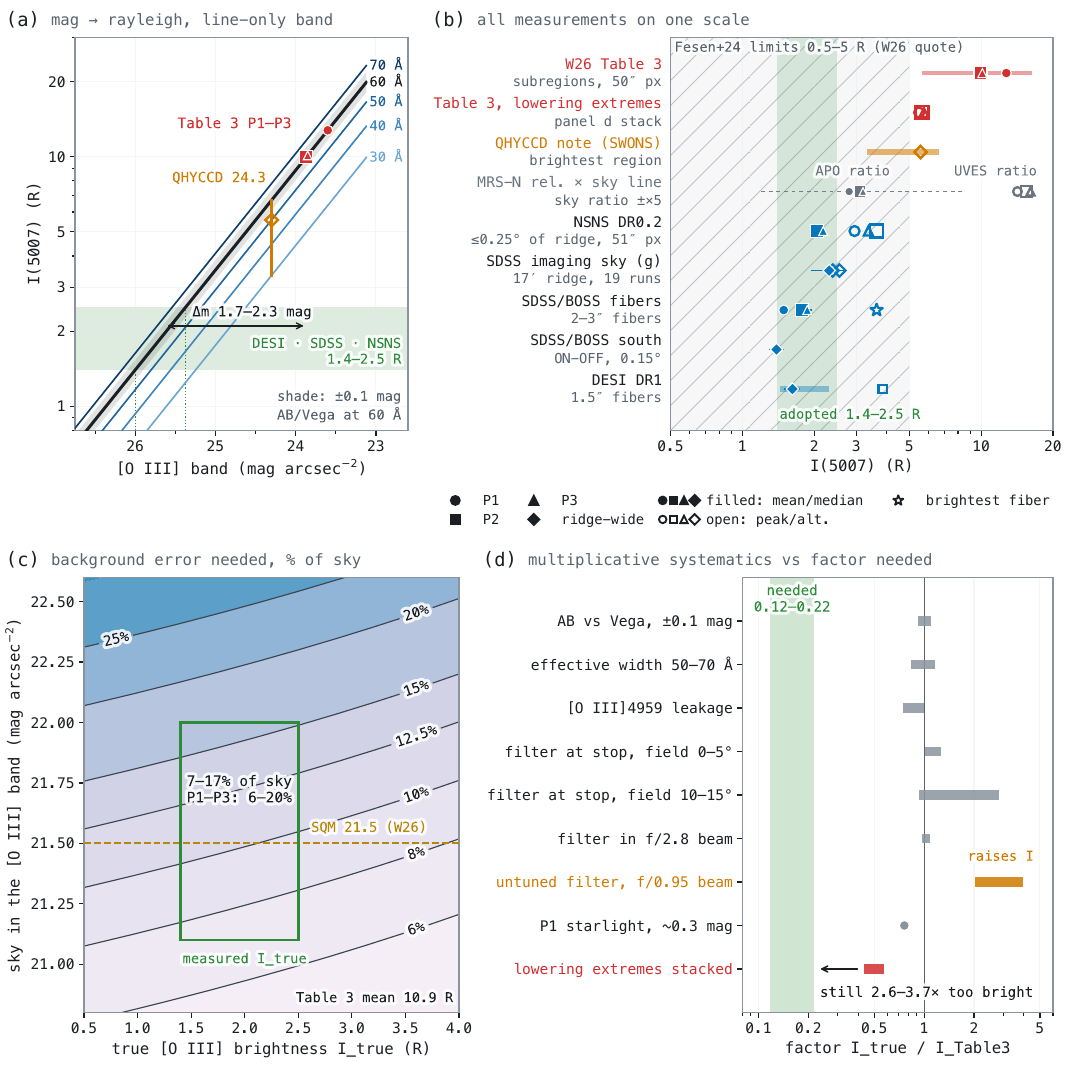}
\caption{\Oiii~$\lambda$5007 surface brightness: W26 \citep{Wu2026} Table~3 and independent data. (a) Conversion of \Oiii-band AB surface brightness to $I(5007)$ for 30--70\,\AA\ line-only bands, with W26 Table~3, the 24.3\,mag\,arcsec$^{-2}$ reported by \citet{QHYCCD2024} for the brightest region (bar, 30--60\,\AA) and the adopted range (band). (b) Measurements by method and angular scale: W26 Table~3 (bar, conversion variants) and its combined lowering extremes; QHYCCD; MRS-N relative intensities scaled by sky-line ratios from SDSS/BOSS at Apache Point Observatory (APO; dashed, 16--84\%) or from the VLT/UVES sky atlas \citep{Hanuschik2003}; and NSNS, SDSS and DESI data \citep{Ziegenbalg2025,Almeida2023,Abdurrouf2022,SDSS2026,DESI2026}. Hatched, the 0.5--5\,R that W26 quotes for \citet{Fesen2024}; green band, adopted 1.4--2.5\,R. (c) Additive error, as a fraction of the in-band sky, that raises $I_{\rm true}$ to the W26 Table~3 mean of 10.9\,R; box, measured $I_{\rm true}$ and plausible skies; dashed line, W26 sky-quality-meter (SQM) reading. (d) Factors $I_{\rm true}/I_{\rm Table\,3}$ from single systematics (generic three-cavity 60-\AA\ filter model) and all lowering extremes combined; band, the required 0.12--0.22. The rows for field angles of 10--15$^\circ$, an f/2.8 beam and an untuned filter in an f/0.95 beam, not used in the text, give 0.93--4.0.}
\label{fig:brightness}
\end{figure*}

Independent data are lower (Figure~\ref{fig:brightness}b). DESI DR1 spectra \citep{DESI2026} give a southern-ridge median of 1.62\,R ($\approx$20\% calibration uncertainty; Appendix~\ref{sec:appendix-desi}); SDSS sky fibers \citep{Abdurrouf2022,SDSS2026} average 1.49, 1.78 and 1.87\,R within 0.3$^\circ$ of P1--P3; SDSS imaging sky levels \citep{Almeida2023} give $2.31\pm0.38$\,R on 17$'$ scales; and NSNS maps \citep{Ziegenbalg2025} give 2.04, 2.06 and 2.18\,R. We adopt 1.4--2.5\,R. Per region, W26 Table~3 is 4.6--8.6 times higher. It exceeds even the brightest single measurements (3.65\,R in an SDSS fiber, 3.88\,R at a DESI node, 2.95--3.64\,R in NSNS $1'$ peaks) by factors of 2.6--4.3, so angular scale does not explain it.

No identified multiplicative systematic reaches the required factor of 0.12--0.22 (Figure~\ref{fig:brightness}d): the magnitude system ($\pm0.1$\,mag; \citealt{Blanton2007}), the band width (50--70\,\AA), \Oiii~$\lambda$4959 leakage and a filter in the lens aperture (generic model; field angles 0--5$^\circ$) individually give $I_{\rm true}/I_{\rm Table\,3}=0.75$--1.27, and all lowering extremes combined, with $\sim$0.3\,mag of P1 starlight, yield 5.5--5.8\,R, still 2.6--3.7 times the independent values. A common factor would also preserve the regional ranking, yet W26 Table~3 ranks P1 brightest, which W26 Sect.~3.1 attributes to starlight, and SDSS ranks it faintest.

An additive error of 7.9--11.3\,R, 6--20\% of an in-band sky of 21.1--22.0\,mag\,arcsec$^{-2}$ (Figure~\ref{fig:brightness}c), would reproduce W26 Table~3; a background or flat-field residual or unequal starlight between source and background apertures could supply it, but the published information does not show which.

For a shock-excited filament, as W26 proposes, published shock models make 10--13\,R inconsistent with the far ultraviolet. GALEX FUV follows the southern filament at $92\pm10$\,CU per NSNS rayleigh (Section~\ref{sec:counterparts}). W26 Table~3 would make this 15--20\,CU\,R$^{-1}$, 2.9--3.8 times below the 57\,CU\,R$^{-1}$ non-resonant FUV floor of the incomplete shocks of \citet{Raymond1988} at $E(B-V)=0.03$ \citep{Schlegel1998}, the lowest among tabulated shock models \citep{Raymond1979,Cox1985,Dopita1996}.

At 1.4--2.5\,R the filament lies within the 0.5--5\,R range that W26 quotes for \citet{Fesen2024}, and its detection by \citet{Fesen2026} does not require a bright filament.

\subsection{Red-line ratios do not classify the \texorpdfstring{\Oiii{}}{[O III]} filament}\label{sec:shock}

W26 Sect.~4.1 classifies the ionized gas as shock-excited from red-line criteria that separate remnants from \ion{H}{2} regions \citep{Fesen1985}: \Sii~$\lambda\lambda$6716,\,6731/\Ha{} ``exceeds 0.5 in all four regions'', \Nii/\Ha{} lies within 0.5--1.5, and the points fall in the remnant zones of the \citet{Sabin2013} and \citet{Kopsacheili2020} diagrams (W26 Fig.~8, whose three panels use only \Ha, \Nii{} and \Sii). The \Sii/\Ha{} threshold varies within W26 between 0.4 and 0.5.

\begin{figure*}
\centering
\includegraphics[width=\textwidth]{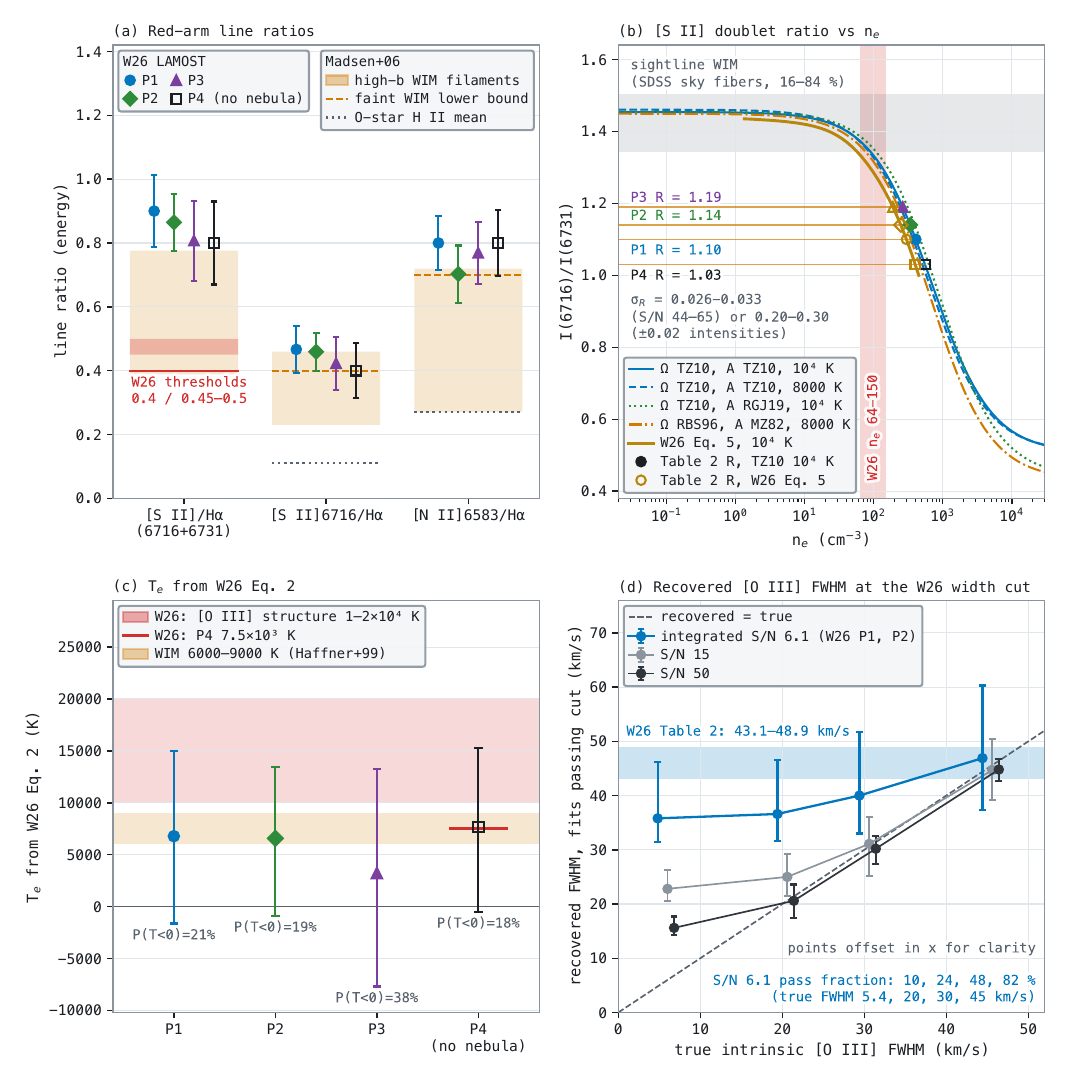}
\caption{Tests of the W26 \citep{Wu2026} diagnostics with W26 Table~2 values (P1--P3, filament; P4, nebula-free). (a) Red-arm line ratios (energy units; $1\sigma$ errors) with the warm-ionized-medium (WIM) and \ion{H}{2}-region values of \citet{Madsen2006}; the shading for the doublet sum assumes the low-density limit. (b) $I(6716)/I(6731)$ versus electron density $n_e$ for a five-level S$^+$ atom (collision strengths $\Omega$ and transition probabilities $A$ from TZ10, \citealt{Tayal2010}; RBS96, \citealt{Ramsbottom1996}; MZ82, \citealt{Mendoza1982}; RGJ19, \citealt{Rynkun2019}) and for W26 Eq.~5 \citep{Proxauf2014}; markers, W26 Table~2 ratios; gray band, WIM in SDSS sky fibers \citep{Abdurrouf2022,SDSS2026}. (c) $T_e$ from W26 Eq.~2 (Monte Carlo 16th--84th percentiles) and $P(T<0)$; WIM range from \citet{Haffner1999}. (d) Recovered versus true \Oiii{} FWHM (median, 16th--84th percentiles) of simulated fits passing the W26 width cut, by integrated S/N.}
\label{fig:redline}
\end{figure*}

These ratios do not distinguish the filament from its surroundings. The W26 Table~2 intensities give \Sii/\Ha{} $=0.90\pm0.11$, $0.86\pm0.09$ and $0.81\pm0.13$ in P1--P3 and $0.80\pm0.13$ in the nebula-free region P4, and \Nii~$\lambda$6583/\Ha{} $=0.80$, 0.70 and 0.77 ($\pm0.09$--0.10) against $0.80\pm0.10$ (Figure~\ref{fig:redline}a); no filament region differs from P4 by more than $0.72\sigma$. P4 therefore meets the W26 criteria too, as W26 Sects.~4.1 and~4.2 note, and in all three panels of W26 Fig.~8 the background points overlap the filament points; W26 nevertheless concludes that the remnant dominates and the diffuse-gas contribution is minor.

The values are those of the faint high-latitude WIM. \citet{Madsen2006} find \Sii~$\lambda$6716/\Ha{} $\gtrsim0.4$ where \Ha{} is faint ($\gtrsim0.68$ for the doublet sum at the low-density limit) and \Nii/\Ha{} $\gtrsim0.7$ in the faintest WIM at $\ell>150\arcdeg$, against means of 0.11 and 0.27 in O-star \ion{H}{2} regions; both ratios rise as \Ha{} fades, which they and \citet{Haffner1999} attribute to WIM temperature variations. Toward P1--P4 the nearest WHAM beams \citep{Haffner2003} give 2.9--4.4\,R of \Ha, and neither WHAM (on$-$off $\Delta I=+0.04\pm0.35$\,R in the south) nor the SDSS sky fibers ($0.08\pm0.22$\,R at Dec $<17.6\arcdeg$) shows an excess on the filament.

The MRS-N red lines therefore trace foreground and background WIM, consistent with the diffuse-ionized-gas contribution inferred in W26 Sect.~4.2; thresholds that separate remnants from \ion{H}{2} regions do not separate them from the WIM at a few rayleighs. With no \Hb{} in the blue arm and no red-line ratio that isolates the \Oiii-emitting gas, the MRS-N data do not constrain the excitation of the filament. This is not evidence against shock excitation, which Section~\ref{sec:excitation} tests with \Oiii/\Hb{} and other diagnostics.

\subsection{Other remnant diagnostics}\label{sec:snrdiag}

\textit{Electron density.} W26 derives a mean $n_e\approx64$\,cm$^{-3}$ from 107 per-fiber values (W26 Fig.~7b), rising to $\approx150$\,cm$^{-3}$ toward the center, from the \Sii{} doublet with its Eq.~5 \citep{Proxauf2014} and assigns it to the \Oiii-emitting gas (W26 Sects.~3.4, 4.1). The regional W26 Table~2 ratios, $R=I(6716)/I(6731)=1.03$--1.19, give 169--400\,cm$^{-3}$ with W26 Eq.~5 and 225--631\,cm$^{-3}$ with a five-level S$^+$ atom at 8000--$10^4$\,K \citep{Tayal2010,Ramsbottom1996,Mendoza1982,Rynkun2019}. They lie 5.6--9.2$\sigma$ below the $R=1.33$ that W26 Eq.~5 requires for 64\,cm$^{-3}$ if $\sigma_R=0.026$--0.033 (W26 Table~2 S/N), and constrain no density if $\sigma_R=0.20$--0.30 (independent $\pm0.02$ intensity errors; Figure~\ref{fig:redline}b). Either way, the doublet traces the WIM, which SDSS sky fibers and DESI place at the low-density limit on this sightline ($R=1.40$; 1.34--1.50 and 1.33--1.49). The \Oiii{} gas needs only $n_e\approx0.1$--3\,cm$^{-3}$ to emit 2--4\,R (measured: 1.4--2.5\,R; Section~\ref{sec:brightness}) at $\approx$280\,pc (Section~\ref{sec:location}).

\textit{Temperature and turbulence.} W26 Eq.~2 \citep{Reynolds1977} and W26 Table~2 \Ha{} and \Nii{} widths give $T=6780$, 6580, 3220 and 7660\,K for P1--P4, with $\sigma_T=7300$--10700\,K and $P(T<0)=18$--38\% (Figure~\ref{fig:redline}c), so the regional widths do not show P1--P3 to be hotter than P4, as W26 Sect.~3.4 reports from per-fiber values (1--2$\times10^4$\,K on the structure, $7.5\times10^3$\,K in P4). W26 Eq.~3 gives $V_t=25.7$--29.9\,\kms{} for P1--P3 and 26.5\,\kms{} for P4 (W26 Table~4: 30.6--33.1 and 29.9\,\kms); P4 again matches P1--P3, as expected for WIM broadening.

\textit{Abundances.} The N2 index, $\log$(\Nii~$\lambda$6583/\Ha) $=-0.153$ to $-0.097$, lies outside the calibrated range of W26 Eq.~6 \citep[$-2.5<{\rm N2}<-0.3$;][]{Pettini2004}. P1--P3 give $12+\log({\rm O/H})=8.81$--8.84 with W26 Eq.~6 and 8.64--8.72 with N2S2H$\alpha$ \citep{Dopita2016}; P4 gives 8.84 and 8.75. Both calibrations assume stellar photoionization, whereas in the WIM a high \Nii/\Ha{} traces a higher electron temperature, not a higher N/H \citep{Haffner1999,Madsen2006}; the derived O/H describes the WIM.

\textit{\Oiii{} line width.} The intrinsic \Oiii{} widths in W26 Table~2 (FWHM 43.1--48.9\,\kms) do not distinguish an unresolved line from a 45\,\kms{} one: simulated fits that pass the acceptance cut of W26 Sect.~3.2.2 (assumed blue-arm instrumental FWHM $40\pm1.5$\,\kms) at S/N 6.1 (W26 Table~2), taken as integrated, return median FWHM of 35.8--46.9\,\kms{} for true FWHM of 5.4--45\,\kms{} (Figure~\ref{fig:redline}d). W26 does not define its S/N; for a peak S/N the bias is smaller and the widths would indicate $\approx45$\,\kms.

\textit{Energetics.} The W26 inputs reproduce the shock velocity $v_s=119.3$\,\kms{} \citep{Dopita1979}, the explosion energy $E=2.01\times10^{51}$\,erg \citep{Boumis2022} and, within the quoted uncertainties, the radius at the onset of the pressure-driven snowplow (PDS) phase ($R_{\rm PDS}=17.1$\,pc against $18\pm2.8$\,pc) and the age ($7.95\times10^4$\,yr against $(6.8\pm1.8)\times10^4$\,yr; \citealt{Cioffi1988}); the ``$+1$'' in the printed age formula, where \citet{Cioffi1988} has ``$+1/3$'', is a misprint that does not affect the quoted age. The inputs are unsupported: the shell radius $R_s=30$\,pc follows from the 160\,pc distance (Section~\ref{sec:distance}), the ambient density $n_0=1.0\pm0.3$\,cm$^{-3}$ is adopted, and $v_s$ derives from the WIM \Sii{} density. A complete radiative shock emits at least 0.117 \Hb{} photons per swept-up H atom, so the \Hb{} non-detection in the spectra of Section~\ref{sec:spectrum} ($I({\rm H}\beta)\le0.15$\,R at $2\sigma$, from \Oiii/\Hb{} $>11$ at 1.62\,R in DESI), with limb brightening of 3--18.6, bounds such a shock at $\approx$280\,pc to $n_0\le0.005$--0.043\,cm$^{-3}$ and $E\le4.2\times10^{49}$--$3.8\times10^{50}$\,erg (model scalings); incomplete shocks (Section~\ref{sec:excitation}) are exempt.

\textit{Radio.} W26 Sect.~4.1 takes the absence of counterparts in the NRAO VLA Sky Survey (NVSS; \citealt{Condon1998}) and the TIFR GMRT Sky Survey (TGSS; \citealt{Intema2017}) as consistent with an old remnant. These interferometric surveys resolve out smooth structure wider than $\approx8\arcmin$ \citep[NVSS snapshots;][]{NRAO2025} and 10\arcmin--20\arcmin{} \citep[TGSS;][]{Intema2017}; the filament is 17\arcmin{} wide, so the non-detection is uninformative.

\section{New constraints from archival survey data}\label{sec:new}

\subsection{The filament spectrum}\label{sec:spectrum}

DESI DR1 alone (Section~\ref{sec:data}) detects the filament from Dec $+14^{\circ}$ to $+28^{\circ}$ at $\approx30\sigma$ against empirical nulls; SDSS sky fibers detect it along its whole length. The \Oiii~$\lambda$5007 ridge is 1.4--2.5\,R in the south (Section~\ref{sec:brightness}) and 0.4--0.9\,R in the north.

The \Oiii\ velocity, \vhel\ $=+13.3\pm1.0$\,\kms\ (Table~\ref{tab:velocity}), is the same on the E and W strands (DESI: $+12.0$ [11.3--14.3] and $+11.4$ [8.6--15.6]\,\kms). North of Dec $+19.5^{\circ}$ the DESI centroid is redder by $+4$ to $+9$\,\kms\ in \vhel\ ($\Delta V_{\rm LSR}=+7$ to $+11$\,\kms), at 1.8--3.7$\sigma$ for the six estimators. SDSS gives a consistent $+6.7\pm6.5$\,\kms\ but disagrees with DESI at Dec 17.6--21$^{\circ}$ ($+10.0\pm2.2$ against $+17$ to $+23$\,\kms; 2--3$\sigma$), so the northern shift rests on DESI, with an estimator-dependent amplitude.

\begin{figure*}
\includegraphics[width=\textwidth]{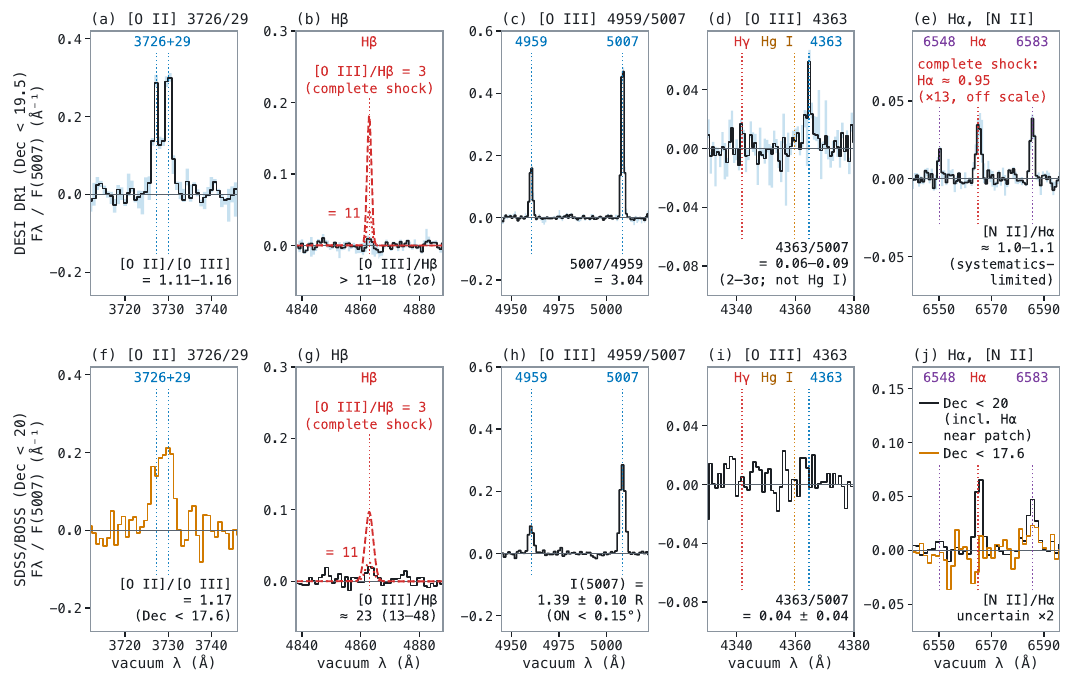}
\caption{Spectrum of the filament: flux density per \AA\ in units of the \Oiii~$\lambda5007$ flux (energy units) against vacuum wavelength. (a)--(e) DESI DR1 template regression \citep{DESI2026} for trace Dec $<19.5^{\circ}$; shading, 16--84\% HEALPix block-bootstrap range. (f)--(j) SDSS/BOSS total-sky ON$-$OFF stacks \citep{Abdurrouf2022} at Dec $<20^{\circ}$ (ON fibers within $0.15^{\circ}$ of a strand), except (f), seven BOSS groups at Dec $<17.6^{\circ}$ normalized to their own \Oiii~$\lambda5007$, and the Dec $<17.6^{\circ}$ curve in (j), which excludes \Ha\ structure near the \Ha\ patch. In (b) and (g), Gaussians at the \Oiii\ line width show \Oiii/\Hb\ $=3$ (dashed; complete shock) and 11 (dotted); (d) and (i) mark H$\gamma$ and \ion{Hg}{1}~$\lambda4358$; (e) notes the \Ha\ of a complete shock (0.95, off scale). Dotted vertical lines mark the lines at \vhel\ $=+13$\,\kms; annotations give measured values (Table~\ref{tab:ratios}).\label{fig:spectra}}
\end{figure*}

The filament is Balmer-dark (\Oiii/\Hb\ $>11$; Table~\ref{tab:ratios}; Figure~\ref{fig:spectra}b,g), with \Oii\ as strong as \Oiii\ in the south and no detected \ion{He}{2}. A 2--3$\sigma$ feature at 4364.55--4364.73\,\AA\ matches \Oiii~$\lambda$4363 at the filament velocity, not \ion{Hg}{1}~$\lambda$4358 (Figure~\ref{fig:spectra}d); if real, it implies $T_e\approx(3$--$6)\times10^4$\,K; \Oii~$\lambda\lambda7320,7330/\lambda3727$ gives $T(\mathrm{O\,II})\approx(2.4$--$5.5)\times10^4$\,K at low density. Northward, \Oii/\Oiii\ rises smoothly from 1.1--1.2 to 4--8 as \Oiii\ fades and \Oii\ brightens (SDSS: 1.9 to 3.0\,R); with \Oiii/\Hb\ $\approx6$--8 there (SDSS central values), the north is of lower excitation. The \Ha\ patch at $(\alpha,\delta)=(115\fdg47,+20\fdg54)$ has a low-ionization SDSS ON$-$OFF spectrum, from which the smooth WIM cancels (\Sii/\Ha\ $=1.04$, \Nii/\Ha\ $=0.78$, \Oii/\Hb\ $\approx13$, \Oiii/\Hb\ $=1.7$) with an unreddened case-B decrement (\Ha/\Hb\ $=2.84$), qualitatively that of a nearly or fully cooled radiative shock in denser gas.

\begin{deluxetable*}{llll}
\tabletypesize{\small}
\tablecaption{Line ratios of the Silk Nebula filament\label{tab:ratios}}
\tablehead{\colhead{Ratio} & \colhead{DESI DR1} & \colhead{SDSS/BOSS} & \colhead{Adopted}}
\startdata
\Oiii~$\lambda5007$/\Hb & $>11$ (box), $>18$ (Gaussian) & $\approx23$ (13--48)\tablenotemark{a} & $>11$; plausibly 15--25 \\
$R_{23}$ & $>29$--34 & $\approx76$ ($>20$) & $\gtrsim30$ \\
\Oii~$\lambda3727$/\Oiii~$\lambda5007$ & 1.11--1.16 (Dec $<19.5^{\circ}$) & 1.17 (Dec $<17.6^{\circ}$) & 1.1--1.2 (south); 4--8 (north) \\
\ion{He}{2}~$\lambda4686$/\Oiii~$\lambda5007$ & $<0.026$ & $<0.04$ & $<0.026$ \\
\Oiii~$\lambda4363/\lambda5007$ & 0.06--0.09 (2--3$\sigma$) & $0.04\pm0.04$ & tentative; $T_e\approx(3$--$6)\times10^4$\,K \\
\Oii~$\lambda\lambda7320,7330/\lambda3727$ & 0.060 (0.055--0.073)\tablenotemark{b} & \nodata & $T(\mathrm{O\,II})\approx(2.4$--$5.5)\times10^4$\,K \\
\Oii~$\lambda3729/\lambda3726$ & $1.31\pm0.07$ & not constraining\tablenotemark{c} & inconclusive \\
\Oiii~$\lambda5007$/\Ha\ (photons) & $>3.4$ & $>3.0$ (Dec $<17.6^{\circ}$) & $\gtrsim3$ \\
\relax[\ion{Ne}{3}]~$\lambda3869$/\Oiii~$\lambda5007$ & 0.07--0.10 & \nodata & 0.07--0.10 (low confidence) \\
\enddata
\tablecomments{Energy ratios unless marked. DESI values are regressions on all filament fibers, dominated by the bright southern ridge, unless a declination range is given; SDSS/BOSS values are total-sky ON$-$OFF stacks at Dec $<20^{\circ}$ unless stated. Limits are $2\sigma$ (\ion{He}{2}: 97.5\%). $R_{23}=$ (\Oii~$\lambda3727$ + \Oiii~$\lambda\lambda4959,5007$)/\Hb. Line-free null windows scatter by 0.02 (rms; largest 0.05) in these ratios, so DESI ratios below $\approx0.05$ are effectively upper limits; the \ion{He}{2} limit is a bootstrap percentile that does not include this scatter. In the last column, south and north refer to Dec $\lesssim17.6^{\circ}$ and $\gtrsim21^{\circ}$.}
\tablenotetext{a}{Photon ratio; $2\sigma$ limits including placebo systematics are $>5$--7.5.}
\tablenotetext{b}{Four-line fit (south: 0.064); $\pm0.015$ systematic from residual OH emission.}
\tablenotetext{c}{The doublet is blended at BOSS resolution, and the blue-end wavelength and line-spread calibrations bias fixed-velocity fits.}
\end{deluxetable*}

\subsection{Counterparts at other wavelengths and precovery}\label{sec:counterparts}

\begin{figure*}
\includegraphics[width=\textwidth]{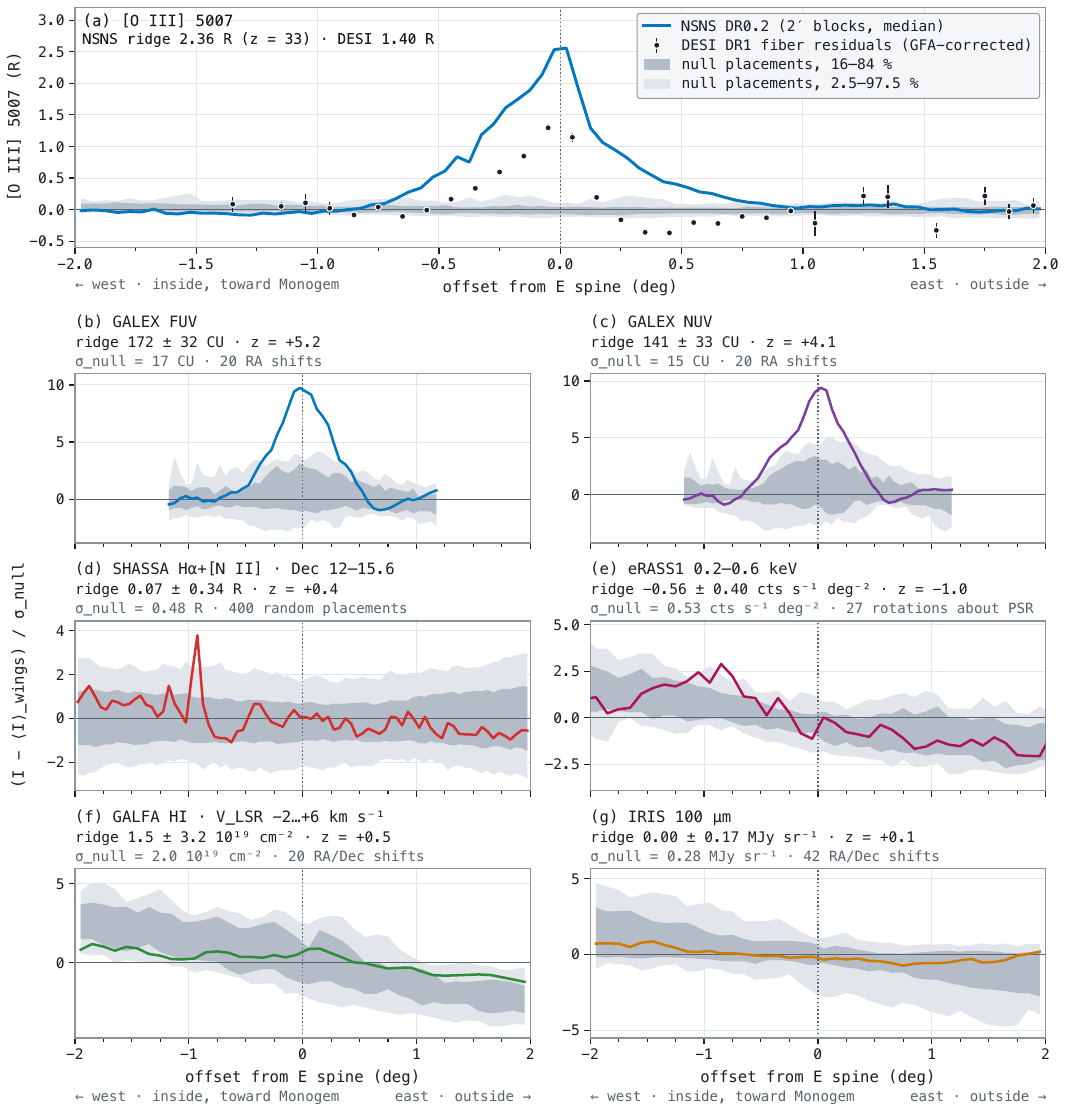}
\caption{Profiles across the southern spine (Dec $<17.6^{\circ}$) against offset $x$ from the E spine (positive eastward). (a) \Oiii~$\lambda$5007: NSNS DR0.2 map \citep[median in $2'$ blocks;][]{Ziegenbalg2025} and DESI DR1 fiber residuals \citep[mean $\pm$ s.e.;][]{DESI2026}, scaled by the guide-camera (GFA) fiber acceptance and relative to the per-petal sky mean, hence negative beside the ridge. (b, c) GALEX FUV and NUV \citep{Martin2005}; (d) SHASSA \Ha+\Nii\ \citep{Gaustad2001}; (e) eRASS1 \citep{Merloni2024}; (f) GALFA-\ion{H}{1} column density \citep{Peek2018}; (g) IRIS \citep{MivilleDeschenes2005}. In (b)--(g) each profile, minus its mean over the wings ($|x|\ge0.5^{\circ}$), is divided by $\sigma_{\rm null}$, the median per-bin scatter of null profiles (shifted or randomly placed traces; eRASS1: rotations about PSR B0656+14). Shading: 16--84\% and 2.5--97.5\% null ranges. Headers: fitted amplitude of a Gaussian ridge (FWHM $0.28^{\circ}$) $\pm$ the scatter of null amplitudes, and significance $z$.\label{fig:profiles}}
\end{figure*}

GALEX \citep{Martin2005,Morrissey2007} FUV and NUV images show a narrow ridge on the southern \Oiii\ spine (Figure~\ref{fig:profiles}b,c), at 68--92 (FUV; $8.7\sigma$ in a template regression on the \Oiii\ map) and 36--57\,CU (NUV) per rayleigh of \Oiii~$\lambda$5007 on the NSNS scale ($\approx85$--115 for the FUV on the DESI scale; no coverage north of Dec $+22^{\circ}$). Dust-scattered light contributes at most 34 (FUV) and 16\,CU (NUV), well below the measured ridge amplitudes of $172\pm32$ and $141\pm33$\,CU (Figure~\ref{fig:profiles}b,c).

No other tracer shows the filament (Figure~\ref{fig:profiles}d--g). In \Ha, SHASSA \citep{Gaustad2001}, the only arcminute-resolution survey of the spine (Dec 12--15.6$^{\circ}$), gives $<0.77$\,R, and WHAM \citep{Haffner2003} shows the same emission on and off the filament (Section~\ref{sec:shock}). GALFA-\ion{H}{1} \citep{Peek2018} shows no \ion{H}{1} filament, wall, cavity or velocity signature ($<1.0$--$1.75\times10^{19}$\,cm$^{-2}$ per 2\,\kms\ channel), IRIS 100\,$\mu$m \citep{MivilleDeschenes2005} no dust ($\Delta E(B-V)<0.004$), eRASS1 0.2--0.6\,keV \citep{Merloni2024} no X-ray filament, and the single-dish 408\,MHz survey \citep{Haslam1982} no total-intensity counterpart: $\Sigma_{\rm 1\,GHz}<1.3\times10^{-22}$\,W\,m$^{-2}$\,Hz$^{-1}$\,sr$^{-1}$ ($2\sigma$; corrected for beam dilution and extrapolated to 1\,GHz with $S_\nu\propto\nu^{-0.5}$).

A ridge template fitted to SDSS imaging sky levels \citep{Almeida2023} in 19 $g$-band runs from 2002--2007 gives $2.3\pm0.4$\,R ($3.9\sigma$ against coherent nulls; 1.4--4.6$\sigma$ across estimators), with no signal in $r$ or $i$, as expected for line emission; the earliest $\ge3\sigma$ run dates from 2002 December~31. SDSS sky fibers detect the filament from 2003 January~30 ($3.4\sigma$) and at $\ge5\sigma$ from 2005 March~3. Its brightness varied by less than a factor of $\approx2$ over 2002--2022. Digitized Sky Survey (DSS) plates do not detect it ($3\sigma$ limit $\approx40$\,R in DSS2 $J$).

\subsection{Excitation mechanism}\label{sec:excitation}

\begin{figure*}
\includegraphics[width=\textwidth]{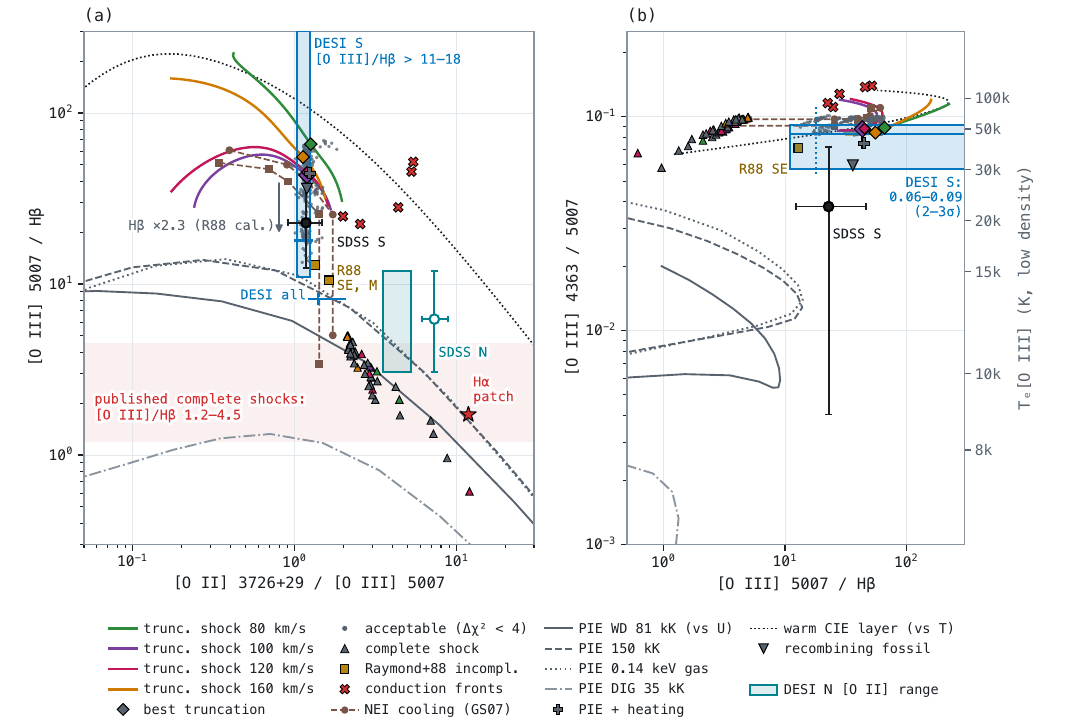}
\caption{Observed line ratios (energy units) and model loci. (a) \Oiii~$\lambda5007$/\Hb\ against \Oii~$\lambda\lambda3726,3729$/\Oiii~$\lambda5007$. (b) \Oiii~$\lambda4363/\lambda5007$ (right axis: low-density $T_e$) against \Oiii~$\lambda5007$/\Hb. S, N: southern, northern filament. DESI~S boxes start at the $2\sigma$ limit \Oiii/\Hb\ $>11$; the tick in (a) and the dotted line in (b) mark the Gaussian-fit limit, 18. SDSS~S combines \Oii/\Oiii\ at Dec $<17.6^{\circ}$ (95\% bar) with \Oiii/\Hb\ at Dec $<20^{\circ}$ (68\% bar). `DESI all' has an arbitrary ordinate; the DESI~N \Oii/\Oiii\ range is drawn over the SDSS~N \Oiii/\Hb\ range. Truncated-shock (trunc.) tracks: cumulative ratios against swept-up column for the best run at each $v_s$; diamonds, best truncations; dots, all runs within $\Delta\chi^2<4$. R88: incomplete-shock models SE and M of \citet{Raymond1988}. PIE: photoionization equilibrium against ionization parameter $U$ (kK: blackbody temperature; WD: WD~J0736+1444; DIG: diffuse ionized gas); plus sign, PIE with extra heating to $4\times10^4$\,K. Inverted triangle: unheated gas recombining from $4\times10^4$\,K. NEI: non-equilibrium cooling of hot gas \citep[GS07;][]{Gnat2007}, isobaric (circles) and isochoric (squares). CIE: collisional ionization equilibrium. Arrow: model \Hb\ $\times2.3$ (R88 calibration). Star in (a): the \Ha\ patch (SDSS; Section~\ref{sec:spectrum}). Shading in (a): published complete shocks \citep{Raymond1979,Cox1985,Dopita1996,Allen2008}.\label{fig:diagnostics}}
\end{figure*}

We compared these constraints with plane-parallel non-equilibrium-ionization models of steady radiative shocks, complete or truncated at a swept-up column $N_*$ (no radiative transfer, so \Hb\ at intermediate columns is about half that of \citealt{Raymond1988} and each fit is repeated with model \Hb\ $\times2.3$; validated against \citealt{Raymond1979,Raymond1988,Allen2008,Gnat2007}; ultraviolet emissivities from CHIANTI v7.1, \citealt{Dere1997,Landi2013}), and with simplified models of the alternatives (Figure~\ref{fig:diagnostics}).

The spectrum rejects photoionization equilibrium at $p\le0.007$ ($\chi^2=21$--51 for 8 terms; $\chi^2=18$--25, $p\le0.02$, if $\lambda4363$ is floor-limited), which reaches only \Oiii/\Hb\ $\lesssim10$--14 and $\lambda4363/\lambda5007\lesssim0.012$--0.034 (diffuse ionized gas: \Oiii/\Hb\ $\le1.3$), and gives \ion{He}{2}/\Oiii\ $=0.06$ for a hot-gas ionizing spectrum (observed: $<0.026$). Even the most favorable cataloged ionizing source, the 81\,kK white dwarf WD~J0736+1444 \citep{Reindl2023}, gives an ionization parameter 18--30 times too low. Complete radiative shocks are excluded ($\chi^2\ge37$ for 10 terms): they reach only \Oiii/\Hb\ $\le5$ (published grids: 1.2--4.5; \citealt{Raymond1979,Cox1985,Dopita1996,Allen2008}), would produce an \Ha\ filament and cooled \ion{H}{1}, and their non-resonant ultraviolet lines alone (85--277\,CU per R) reach or exceed the measured FUV.

Turbulent mixing layers \citep{Slavin1993} and non-equilibrium cooling of the Monogem hot gas \citep{Gnat2007} can match the optical ratios but predict a soft X-ray filament $10^3$--$10^4$ times brighter than the Monogem shell, which eRASS1 excludes. Conduction fronts \citep{Borkowski1990} match them marginally but are excluded by their non-resonant FUV floor of 179--213\,CU per R, above the measured 68--92; they would also need $P/k\approx9\times10^5$\,K\,cm$^{-3}$ to be as narrow as the filament (1.4\,pc at 282\,pc; Section~\ref{sec:location}). A layer statically heated to $\approx6\times10^4$\,K fits the optical ratios ($\chi^2=6.2$ for 9 terms) but requires an unidentified, shock-sized heating flux ($0.5$--$2.6\times10^{-5}$\,erg\,cm$^{-2}$\,s$^{-1}$), and predicts \ion{C}{4}/\Oiii\ 4--10 times below the 0.15--0.4 (photons) implied by the FIMS/SPEAR map of \citet{Kim2007} if its region R1 (a $\approx2\sigma$ spatial match) is the Silk; it is disfavored but not excluded. Photoionization with extra heating to $4\times10^4$\,K also fits the optical ratios ($\chi^2=1.0$--2.4 for 9 terms) but needs 78--350 times the photoionization heating rate and the same missing ionizing source.

A column-truncated (incomplete) radiative shock \citep{Raymond1988} reproduces the line ratios ($\chi^2=0.3$ for 10 terms and four parameters, with 279 of the 480 grid models within $\Delta\chi^2=4$; Figure~\ref{fig:diagnostics}) and the FUV level, but overpredicts the NUV by a factor of 1.5--2 (79--99 against 36--57\,CU per R), comparable to the model systematic. In this model the post-shock gas radiates \Oiii\ and \Oii\ as it cools, but the flow ends before hydrogen recombines, leaving the Balmer lines weak. The optical ratios require $v_s\ge80$\,\kms\ and $\log N_*=17.4$--18.7 (cm$^{-2}$); the FUV level favors 80--120\,\kms\ ($\lesssim150$\,\kms). The acceptable models predict \Oiii/\Hb\ $=13$--62, \Oii/\Oiii\ $=1.0$--1.8 and $\lambda4363/\lambda5007=0.08$--0.10. The nearly symmetric southern cross-profile ($\sigma_W/\sigma_E=1.07$) favors an edge-on flat or gently curved sheet over a smooth curved limb ($\approx2\sigma$). For such a sheet at 282\,pc the models give $P/k\approx1.2\times10^4$--$5.5\times10^5$\,K\,cm$^{-3}$ and a pre-shock density $n_0\approx0.012$--0.13\,cm$^{-3}$ (both $\propto d^{-1}$), the latter $\sim$35--370 times lower than in the Cygnus Loop filaments of \citet{Raymond1988}; resonance scattering of \ion{C}{4} within the sheet brings the predicted FUV to 68--143\,CU per R.

Two tensions remain. DESI places the \Oii\ ridge 0.04--0.05$^{\circ}$ east of the \Oiii\ ridge, whereas a plane shock places it downstream, to the west (disfavored at $p\approx0.003$--0.07). The DESI \Oii~$\lambda3729/\lambda3726$ ratio (Table~\ref{tab:ratios}) lies 2.1--2.7$\sigma$ below the low-density limit that every scenario predicts.

\subsection{Location and distance}\label{sec:location}

An edge-on shock moves in the plane of the sky, so its line-of-sight velocity is that of the pre-shock gas. The \Oiii\ shares the velocity of the local WIM and \ion{H}{1} (\vlsr\ $\approx0$--2\,\kms); the northern redshift matches a northern \ion{H}{1} component at \vlsr\ $=+8.5$ to $+11.3$\,\kms, possibly the pre-shock gas.

The southern spine (Dec $<17.6^{\circ}$) lies $12.10\pm0.12^{\circ}$ from the two-remnant center of \citet{Knies2024}, close to their Monogem shell radius ($11.0\pm1.1^{\circ}$), and follows the outer half-maximum contour of the eRASS1 0.2--0.6\,keV emission in its position-angle sector for every proposed center (offsets $-0.06\pm0.14^{\circ}$ to $+0.19\pm0.09^{\circ}$; $p\approx0.03$--0.05 against rotations). A rigid trace placed at random in the same latitude band and reaching the rim matches the shell radius about any of three proposed centers with probability 0.09. These coincidences favor, but do not establish, a location on the Monogem limb, which lies at 252--318\,pc (68\%; median 282\,pc) if the shell is centered at the parallax distance of PSR B0656+14 ($288^{+33}_{-27}$\,pc; \citealt{Brisken2003}). The southern spine is straighter than this circle (radius of curvature $\approx67^{\circ}$) and only tangent to it; north of Dec $+18.5^{\circ}$ the arc leaves it.

Two further tests do not strengthen the association. Stellar extinctions \citep{Anders2022,Gaia2023} suggest that the cold clump PGCC G199.09+19.95 \citep{Planck2016}, which adjoins the \Ha\ patch on the same circle, lies beyond the limb, at $\approx400$--650\,pc (2--2.5$\sigma$; catalog value $262\pm121$\,pc). E1, a fainter \Oiii\ filament farther east \citep{Fesen2026}, has a similar Balmer-faint DESI spectrum but \vlsr\ $=-24\pm2.5$\,\kms, so the eastern Monogem filaments do not share one velocity.

No measurement fixes the distance: the velocity carries no distance information (Section~\ref{sec:distance}), no absorption line brackets the ionized gas, and no dust is detected. Distances of $\sim$100\,pc to $\sim$1\,kpc are allowed, including the 160\,pc adopted by W26. The $\approx1.6$\,kpc implied by W26 Table~4 is disfavored: the filament would be $\approx470$\,pc long at a height $z\approx0.55$\,kpc above the Galactic plane, with a north--south velocity gradient opposite in sign to Galactic rotation.

\section{Decisive observations}\label{sec:future}

Two questions remain open: whether the filament is a moving shock or a statically heated layer, and how distant it is. Table~\ref{tab:predictions} compares the measurements with the scenario predictions of Section~\ref{sec:excitation}. In order of priority, the observations that address these questions are:

\begin{table*}[t]
\caption{Discriminating observables: measured values and model predictions\label{tab:predictions}}
\footnotesize
\setlength{\tabcolsep}{4pt}
{\centering
\begin{tabular}{llllll}
\hline\hline
Observable & Observed & Truncated shock & Warm layer & Conduction & Photoionized \\
 & & (flat sheet) & (PIE + heating) & front & gas \\
\hline
\ion{C}{4}~$\lambda1548/\lambda1551$ & \nodata & 1.0--1.3 & $\approx2$ & \nodata & \nodata \\
\ion{C}{4}/\Oiii{} (photons) & 0.15--0.4\tablenotemark{a} & 0.04--0.18 & 0.03--0.04 & 1.6--16 & $\approx0$ \\
\Oiii{}/\Hb{} & $>11$ ($2\sigma$) & 17--61 & $\approx97$ (45) & 22--52 & $\lesssim10$--14\tablenotemark{b} \\
\Oiii~$\lambda4363/\lambda5007$ & 0.06--0.09 & 0.082--0.087 & $\approx0.098$ (0.075) & 0.11--0.14 & $\lesssim0.012$--0.034 \\
\Oii~$\lambda7325/\lambda3727$ & 0.060 (0.055--0.073)\tablenotemark{c} & 0.056--0.067 & $\approx0.082$ (0.076) & $\approx0.07$ & \nodata \\
\Oii~$\lambda3729/\lambda3726$ & $1.31\pm0.07$ & \multicolumn{4}{c}{1.455--1.50 in every scenario\tablenotemark{d}} \\
\Oiii{} $\mu$ at 282\,pc (mas\,yr$^{-1}$) & \nodata & 60--120\tablenotemark{e} & $\approx0$ & \nodata & $\approx0$ \\
\Oiii{} FWHM (\kms) & \nodata & 11--13; 20--50\tablenotemark{f} & $\approx13$ & \nodata & $\approx6$ \\
\hline
\end{tabular}\par}
\tablecomments{Energy ratios unless marked; $\lambda7325$ denotes \Oii~$\lambda\lambda7320,7330$. Observed values are DESI DR1 measurements (Table~\ref{tab:ratios}) unless noted; $\lambda4363/\lambda5007$ is a 2--3$\sigma$ measurement. Predictions are model results. Truncated shock: the flat-sheet subset of the acceptable edge-on, column-truncated radiative shock models of Section~\ref{sec:excitation}. Warm layer: collisional-equilibrium layer held at $6\times10^{4}$\,K by unspecified heating; in parentheses, photoionization equilibrium (PIE) plus extra heating to $4\times10^{4}$\,K. Conduction front: static and evaporating planar fronts. Photoionized gas: PIE for hot-star and hot-gas spectra. ``\nodata'': no measurement or no prediction.}
\tablenotetext{a}{Implied by our digitization of the FIMS/SPEAR \ion{C}{4} map of \citet{Kim2007} if the whole excess of its region R1 arises in the Silk (an a~posteriori, $\sim$2$\sigma$ spatial association).}
\tablenotetext{b}{Diffuse ionized gas reaches $\le1.3$.}
\tablenotetext{c}{Plus a $\pm0.015$ systematic from OH residuals.}
\tablenotetext{d}{Low-density limit; the range reflects the atomic data. This row tests the measurement, not the scenarios.}
\tablenotetext{e}{Outward proper motion $\mu=v_s/(4.74\,d)$ for all acceptable flat-sheet models; 60--90 for $v_s=80$--120\,\kms.}
\tablenotetext{f}{Second range: rippled sheet.}
\end{table*}

\begin{enumerate}
\item \textit{Far-ultraviolet spectroscopy} of the southern ridge, from the FIMS/SPEAR archive \citep{Kim2007} or a future mission, separates a shock from static gas independently of the distance. \ion{C}{4} is optically thick in the truncated sheet, so $\lambda1548/\lambda1551=1.0$--1.3, against $\approx2$ for an optically thin warm layer. The \ion{C}{4}/\Oiii{} ratio also differs among the scenarios (Table~\ref{tab:predictions}).

\item \textit{Deep optical spectroscopy} ($R\approx4000$--10000) of the bright southern strand. \Hb/\Oiii{} measured to $\pm0.005$ separates the truncated shock (\Oiii/\Hb{} $=17$--61) from the warm layer ($\approx97$), but not from photoionization plus heating ($\approx45$). \Oiii~$\lambda4363/\lambda5007$ and \Oii~$\lambda\lambda7320,7330/\lambda3727$, the latter to $\pm0.007$ with OH-clean sky subtraction, measure the \Oiii{} and \Oii{} temperatures. The resolved \Oii~$\lambda\lambda3726,3729$ doublet tests the 2.1--2.7$\sigma$ density tension (Section~\ref{sec:excitation}) but cannot rank the scenarios, which all predict the low-density limit.

\item \textit{Two-epoch \Oiii{} proper motion.} At 282\,pc a shock of $v_s=80$--120\,\kms{} moves outward by 60--90\,mas\,yr$^{-1}$; the motion is $\sim$1.8 times larger at 160\,pc, $\sim$3.5 times smaller at 1\,kpc, and zero for a static layer. The SWONS follow-up frames ($\sim$1\farcs55\,pixel$^{-1}$; W26 Sect.~2.1), which date from the discovery period (2023--2024), can provide a first epoch, supplemented by imaging at $\le1''$\,pixel$^{-1}$ in 2026--2027. A second epoch in 2034--2036 detects the motion at 282\,pc if the per-epoch edge precision reaches $\lesssim0\farcs1$--$0\farcs15$. With $v_s$ from the \Hb{} depth or the far-ultraviolet spectrum, the motion gives the distance, $d=v_s/(4.74\,\mu)$.

\item \textit{High-ion absorption bracket.} HST/COS \citep{Green2012} \ion{C}{4} and \ion{Si}{4} absorption toward hot stars at known distances can bracket the filament distance; a Monogem-limb location predicts \ion{C}{4} at the Silk velocity only beyond $\sim$250--320\,pc. From Gaia DR3 \citep{Gaia2023} we select 52 hot stars at 120--800\,pc within $0\fdg3$ of a strand; none has an archival ultraviolet spectrum. By our estimate, a southern bracket of white dwarfs at 234, 294, 445 and 594\,pc costs about 10 orbits. The archival COS G130M spectrum of the background AGN 2MASS~J07452815+1919525, $0\fdg34$ from the W strand, covers \ion{Si}{4} and \ion{N}{5} (not \ion{C}{4}) and is an immediate test for high ions at the Silk velocity.

\item \textit{Spectroscopy of the southern Monogem limb.} Its \Oiii{} arc \citep[here S1;][]{Fesen2026}, e.g.\ at $(\alpha,\delta)=(106\fdg0,+3\fdg47)$ or $(112\fdg5,+4\fdg05)$, lies on the same circle about the two-remnant center of \citet{Knies2024} as the southern Silk. If the southern Silk is the Monogem limb, S1 should show a similar Balmer-dark spectrum at \vlsr{} $\approx0\pm5$\,\kms; a different driver imposes no such requirement. No public DESI or SDSS spectrum constrains S1.

\item \textit{An \Oiii{} profile at $R\ge30000$} tests the geometry rather than the excitation: a flat edge-on sheet gives a thermal FWHM of 11--13\,\kms{} and a static warm layer 13\,\kms, whereas a rippled sheet gives 20--50\,\kms.

\item \textit{Data release.} The per-fiber MRS-N spectra of the W26 plate \citep[not public;][]{Zhang2025}, with their sky templates and wavelength-frame metadata, would allow per-fiber checks of the velocity corrections of Sections~\ref{sec:oiii} and~\ref{sec:halpha}; the SWONS photometric inputs (filter curve, apertures, zero points) would allow a check of Section~\ref{sec:brightness}.
\end{enumerate}

\section{Conclusions}\label{sec:conclusions}

\begin{enumerate}
\item \textit{What stands.} The \Oiii{} filament is real line emission, confirmed by DESI DR1, SDSS and NSNS, and absent from P4. The red-arm lines trace foreground and background WIM, and their \Nii{} and \Sii{} velocities agree with WHAM.

\item \textit{Corrected quantities.} The W26 \Oiii{} velocities require a correction of $-10.1\pm1.0$\,\kms; the combined \vhel{} $=+13.3\pm1.0$\,\kms{} (\vlsr{} $=+1.2\pm1.0$\,\kms) agrees with the red-line, WHAM and local \ion{H}{1} velocities, so the two-layer shell is not supported. The W26 \Ha{} velocities are red-biased by $+4.5$ to $+6.7$\,\kms. The W26 Table~4 distances are in kpc (fiducial $\sim$1.6\,kpc), and kinematics carry no distance information in this direction. The southern \Oiii{} brightness is 1.4--2.5\,R, a factor of 4.6--8.6 below W26 Table~3. The red-line ratios are the same in P4 and describe the WIM. Taken at face value, the W26 Table~2 \Sii{} ratios correspond to 225--631\,cm$^{-3}$, not 64--150\,cm$^{-3}$, but within their tabulated intensity errors they constrain no density, and independent spectra place the WIM doublet at the low-density limit. Like the temperatures, turbulence, abundances and radio non-detection, they do not constrain the filament; the \Oiii{} widths are not established unless the tabulated S/N is a peak value, and the energy and age rest on unsupported inputs.

\item \textit{New constraints.} The filament spectrum is Balmer-dark (\Oiii/\Hb{} $>11$, $2\sigma$) and \Oii-strong (\Oii/\Oiii{} $\approx1.1$--1.2 in the south, 4--8 in the north), with \ion{He}{2}/\Oiii{} $<0.026$ and a 2--3$\sigma$ indication of $T_e\approx(3$--$6)\times10^{4}$\,K. GALEX shows an ultraviolet ridge on the southern spine; no \Ha, \ion{H}{1}, dust or X-ray counterpart is detected. SDSS data show the filament from 2002.

\item \textit{Physical picture.} The spectrum excludes complete radiative shocks and, with the ionizing-photon budget, photoionization equilibrium; X-ray and ultraviolet data exclude mixing layers, hot-gas cooling and conduction fronts; a statically heated warm layer is disfavored but not excluded. A column-truncated radiative shock of $\sim$80--120\,\kms, seen edge-on as a flat or gently curved sheet, reproduces the line ratios and the FUV level; the \Oii--\Oiii{} ridge offset, the \Oii{} doublet ratio and the NUV level remain in tension with it (Section~\ref{sec:excitation}). The southern spine lies at the Monogem shell radius (chance-coincidence probability 0.09; Section~\ref{sec:location}) and on the outer edge of its soft X-ray emission, so the Monogem limb at 252--318\,pc (68\%) is a plausible location. The distance is not measured.

\item \textit{Decisive observations.} Far-ultraviolet and deep optical spectroscopy can separate a shock from a static layer; a two-epoch proper motion and bracketing high-ion absorption can measure the distance (Section~\ref{sec:future}).
\end{enumerate}

\expandafter\let\csname internallinenumbers\endcsname\relax
\begin{acknowledgments}
This work uses the NSNS DR0.2 [\ion{O}{3}] survey of S.~Ziegenbalg. Funding for the Sloan Digital Sky Survey has been provided by the Alfred P. Sloan Foundation and the participating institutions (\url{https://www.sdss.org}). This research uses data obtained with the Dark Energy Spectroscopic Instrument (DESI); DESI is supported by the U.S. Department of Energy Office of Science and the participating institutions (\url{https://www.desi.lbl.gov/collaborating-institutions}). Guoshoujing Telescope (LAMOST) is a National Major Scientific Project built by the Chinese Academy of Sciences. This work is based on data from eROSITA, the soft X-ray instrument aboard SRG. The Wisconsin H-Alpha Mapper and its H$\alpha$ Sky Survey have been funded primarily by the National Science Foundation. This research used the NASA Astrophysics Data System, MAST, IRSA, the NOIRLab Astro Data Lab, SkyView and VizieR.

The author used the AI models Claude Fable~5.2 and Claude Opus~5.5 (Anthropic) for data analysis, programming, figure preparation and drafting of the text. The author takes full responsibility for the results and the content of this paper.
\end{acknowledgments}

\facilities{Sloan, Mayall (DESI), LAMOST, GALEX, eROSITA, WHAM, Arecibo, IRAS, Gaia, Planck}

\software{Astropy \citep{Astropy2022}, NumPy \citep{Harris2020}, SciPy \citep{Virtanen2020}, Matplotlib \citep{Hunter2007}}

\appendix
\restartappendixnumbering

\section{Detection and line fitting in individual DESI spectra near P1 and P3}\label{sec:appendix-desi}

This appendix describes the DESI measurements within $0.5^\circ$ of the W26 regions P1 $(\alpha,\delta)=(116\fdg65,+16\fdg60)$ and P3 $(116\fdg45,+14\fdg72)$ (Figure~\ref{fig:desi_fibers}; Table~\ref{tab:desi_regions}).

\subsection{Spectra}

\begin{figure*}
\centering
\includegraphics[width=\textwidth]{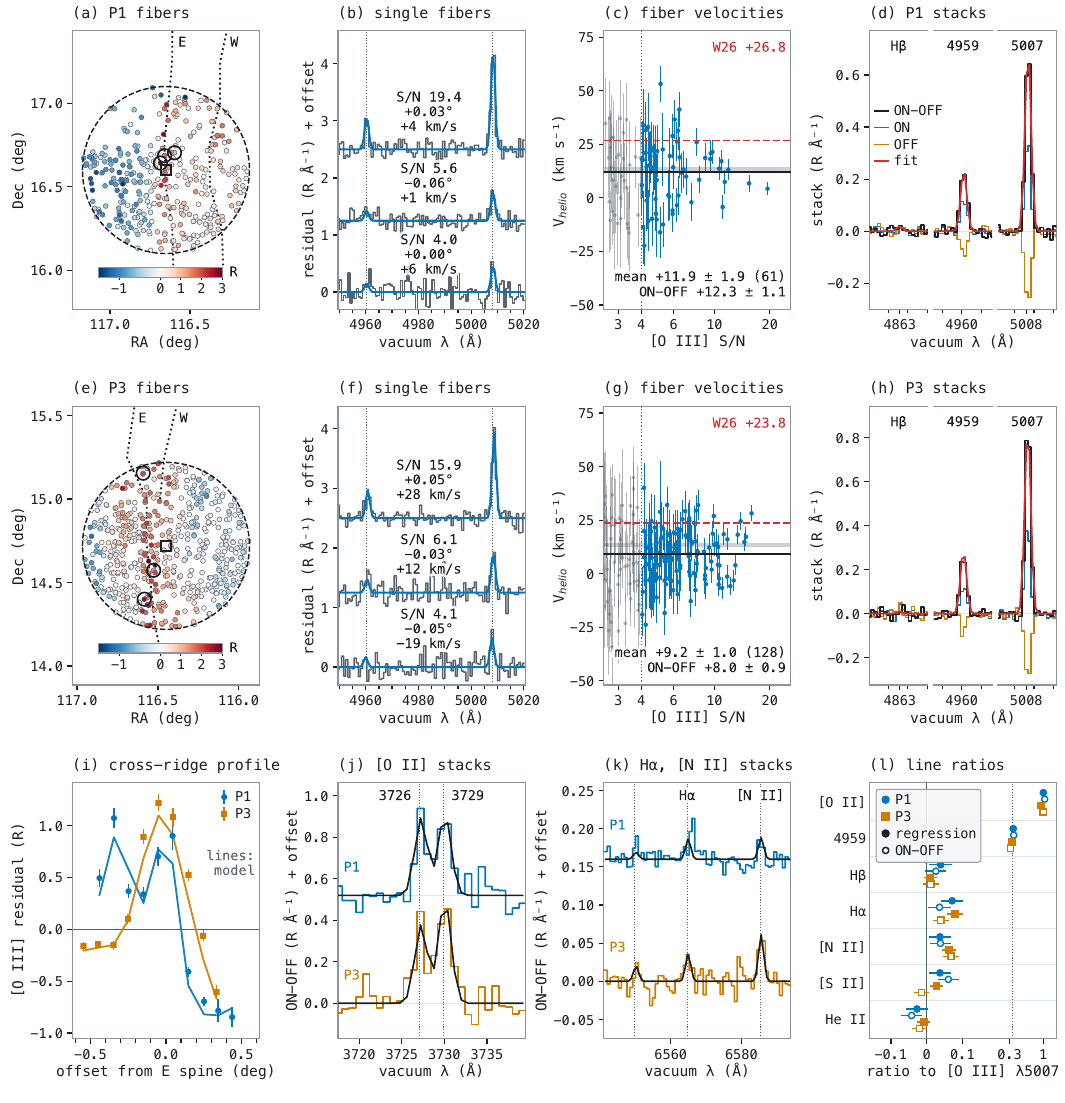}
\caption{DESI DR1 coadds \citep{DESI2026} within $0.5^\circ$ of P1 (top row) and P3 (middle row). (a, e) Fibers with a good fit, colored by petal-relative \Oiii~$\lambda$5007 brightness; dotted lines, E and W strands; dashed circle, $0.5^\circ$ radius; square, region center; rings, the fibers in (b, f). (b, f) Continuum-subtracted residual spectra of the fibers of highest, median and lowest S/N among those with S/N $>4$, with doublet fits; labels give S/N, offset from the E spine and \vhel. (c, g) Fiber \vhel{} against S/N: blue, S/N $>4$; gray, $2.5<$ S/N $\le4$; black line, weighted mean of the blue points; band, adopted value; dashed, W26 Table~4. (d, h) Stacks at \Hb{} and \Oiii~$\lambda\lambda$4959, 5007: ON$-$OFF with fit, ON and OFF. Bottom row, P1 (blue) and P3 (orange): (i) binned residual across the E spine (positive east), with the forward-model prediction (lines); (j, k) ON$-$OFF stacks at \Oii{} and at \Ha{} and \Nii, with fits; (l) ratios to \Oiii~$\lambda$5007 from the regression with intercept (filled) and the ON$-$OFF stack (open); dotted line, 1/2.98.}
\label{fig:desi_fibers}
\end{figure*}

The inputs are not DESI sky-fiber spectra: they are public DR1 coadded spectra \citep{DESI2026} of galaxy and quasar targets with no redshift warning, in $\pm40$\,\AA{} line windows (0.8\,\AA{} pixels; vacuum, barycentric). Stars are excluded because their model mismatch falls on the nebular lines. From each coadd we subtract the best-fit model of the pipeline redshift fitter \citep{Guy2023}. The residuals are relative to the petal sky model (Section~\ref{sec:data}): positive on the ridge, negative beside it (Figure~\ref{fig:desi_fibers}a,e).

We analyze 330 (P1) and 530 (P3) coadds, 32\% and 99.8\% of those eligible (retrieval near P1 was incomplete). Two overlapping survey-validation tiles, 80644 (bright program; six nights) and 80683 (dark; four other nights), supply 280 of the P1 coadds and all P3 coadds; the other 50 are main-survey coadds.

\begin{deluxetable}{lcc}
\tabletypesize{\small}
\tablecaption{DESI DR1 measurements within $0.5^\circ$ of P1 and P3\label{tab:desi_regions}}
\tablehead{\colhead{Quantity} & \colhead{P1} & \colhead{P3}}
\startdata
Coadds: eligible / analyzed / good fit & 1,034 / 330 / 326 & 531 / 530 / 505 \\
Fibers with S/N $>4$ (S/N $<-4$) & 61 (39) & 128 (11) \\
Median per-fiber error (R) & 0.21 & 0.20 \\
Fibers in ON$-$OFF stack: ON, OFF & 161, 175 & 262, 157 \\
\vhel, weighted mean of S/N $>4$ fibers\tablenotemark{a} & $+11.9\pm1.9$ & $+9.2\pm1.0$ \\
\vhel, median of the same fibers & $+13.7\pm2.2$ & $+8.5\pm1.4$ \\
\vhel, ON$-$OFF stack & $+12.3\pm1.1$ & $+8.0\pm0.9$ \\
Model sky brightness at region center (R) & $2.07\pm0.06$ & $1.54\pm0.04$ \\
E-ridge peak (R)\tablenotemark{b} & $1.82\pm0.06$ / $1.83\pm0.06$ / $1.69\pm0.07$ & $1.68\pm0.05$ / $1.65\pm0.03$ / $1.34\pm0.05$ \\
ON$-$OFF mean brightness (R) & $1.36\pm0.05$ & $1.65\pm0.05$ \\
\Oiii~$\lambda5007/\lambda4959$\tablenotemark{c} & 2.91 / 2.97 / 2.87 ($\pm$0.15--0.18) & 3.12 / 3.08 / 3.29 ($\pm$0.17--0.19) \\
\Oii~$\lambda3727$/\Oiii~$\lambda5007$\tablenotemark{c} & 1.00 / 1.03 / 1.05 ($\pm$0.13--0.14) & 0.92 / 0.89 / 0.99 ($\pm$0.09--0.11) \\
\Oii~$\lambda3729/\lambda3726$\tablenotemark{c} & 1.16 / 1.07 / 1.05 ($\pm$0.19--0.24) & 1.28 / 1.26 / 1.32 ($\pm$0.22--0.24) \\
\Oiii~$\lambda5007$/\Hb\tablenotemark{c} & $>9.8$ / $>8.5$ / $>12.0$ & $>21.3$ / $>22.2$ / $>18.1$ \\
\ion{He}{2}~$\lambda4686$/\Oiii~$\lambda5007$ & $<0.08$ & $<0.045$ \\
\Ha/\Oiii~$\lambda5007$ & $<0.16$ & $<0.13$ \\
\Nii~$\lambda6583$/\Oiii~$\lambda5007$ & $<0.13$ & $<0.11$ \\
\Sii~$\lambda\lambda6716,6731$/\Oiii~$\lambda5007$ & $<0.12$ & $<0.06$ \\
\enddata
\tablecomments{Velocities are in \kms{} and exclude the $\pm1.5$\,\kms{} zero-point uncertainty, which these two regions cannot test. Brightnesses use Equation~(\ref{eq:desi_sb}) and exclude the $\approx$20\% acceptance uncertainty of these regions. Ratios are energy ratios; limits are $2\sigma$ and, where one value is given, the weakest of the three estimators. Red-line limits apply to emission that follows the \Oiii{} ridge.}
\tablenotetext{a}{The estimator of Table~\ref{tab:velocity}.}
\tablenotetext{b}{Filament-wide forward model / local fit with the petal term from that model / local fit with a free constant per tile--petal. Uncertainties are bootstraps over fibers.}
\tablenotetext{c}{Regression with intercept / regression with tile--petal means removed / ON$-$OFF stack.}
\end{deluxetable}

\subsection{Line fitting}

For each residual we fit, over 4930--5038\,\AA{} by linear least squares with the pipeline inverse variances, a quadratic continuum plus $A\,[G_{5007}+G_{4959}/2.98]$ (Figure~\ref{fig:desi_fibers}b,f). Here $G$ is a unit-integral Gaussian of fixed dispersion $\sigma=0.759$\,\AA{} (45.5\,\kms) at 5008\,\AA, the quadrature sum of the line-spread function (0.695\,\AA; median of 430 spectra), a 12\,\kms{} intrinsic dispersion and the pixel term. We mask sky lines ($\pm2.5$\,\AA) and the target's own redshifted lines ($\pm8$\,\AA), and clip once at $5\sigma$. The amplitude $A$ at fixed \vhel{} $=+14$\,\kms{} is the $\lambda$5007 flux. The velocity minimizes $\chi^2$ of the same fit on a 2\,\kms{} grid from $-120$ to $+150$\,\kms{} (vacuum rest wavelengths 5008.240 and 4960.295\,\AA); its error follows from the curvature. Other lines are fitted likewise at $+14$\,\kms; the $\lambda5007/\lambda4959$ ratio uses separate fits to the two lines.

Fits pass if at least 60\% of the window is unmasked, the error is below four times the sample median and $|{\rm S/N}|<30$: 326 coadds near P1 and 505 near P3. Of these, 61 and 128 have S/N $>4$ and 39 and 11 have S/N $<-4$; Gaussian noise would give $<0.02$ of each sign. In stacks the line is unresolved (intrinsic FWHM $<40$\,\kms, 97.5\%).

\subsection{Velocities}

The DESI entries of Table~\ref{tab:velocity} are means of the fibers with S/N $>4$ and velocity error $\sigma_V<30$\,\kms, weighted by $1/(\sigma_V^2+5^2)$; the quoted error is the larger of the formal error (1.35 and 0.87\,\kms) and the standard error from the scatter. The fibers scatter more than their formal errors (robust scatter 14.0 and 12.9\,\kms{} against median errors of 10.6 and 9.3\,\kms; Figure~\ref{fig:desi_fibers}c,g), whereas simulated noise gives pulls of rms 1.03--1.05. One cause is that the centroid of a petal-relative residual amplifies velocity differences from the petal mean, here by $\approx$1.9 (P1) and 2.2 (P3).

The petal mean cancels in ON$-$OFF stacks (Figure~\ref{fig:desi_fibers}d,h), formed per tile--petal from fibers with ridge template $T\ge0.5$ (ON) and $T<0.05$ (OFF; anywhere on the petal) and combined with inverse-variance weights; $T$ is the sum of unit-peak Gaussian cross-sections of the traced strands. They give $+12.3\pm1.1$ (P1) and $+8.0\pm0.9$\,\kms{} (P3; bootstrap over fibers), within $1\sigma$ of Table~\ref{tab:velocity}. S/N cuts of 3--8 change the means by $\le1.7$\,\kms, and the two survey-validation tiles differ by $+2.9\pm2.9$ (P1) and $-1.5\pm2.1$\,\kms{} (P3). Within P3, ON$-$OFF gives $+11.2\pm1.1$\,\kms{} east and $+5.5\pm1.0$\,\kms{} west of the E spine (3--4$\sigma$). P1 agrees with the adopted $+13.3\pm1.0$\,\kms; P3 lies 4--5\,\kms{} lower.

\subsection{Line strengths}

DESI fluxes are calibrated for point sources, so a uniform source of surface brightness $I$ yields $F=I\,\Omega/f$, and
\begin{equation}
I=\frac{F f}{\Omega},\qquad I\,[{\rm R}]=0.7428\,f\,F\,[10^{-17}\,{\rm erg\,cm^{-2}\,s^{-1}}]
\label{eq:desi_sb}
\end{equation}
at 5008\,\AA, where $\Omega=\pi(0\farcs76)^2=1.815$\,arcsec$^2$ is the fiber solid angle and $f$ the point-source fiber acceptance (Section~\ref{sec:data}) at the line wavelength.

The predicted residual of fiber $i$ is proportional to $Z_i=T_i-\langle T\rangle$, the mean being over all fiber positions of its tile--petal. The filament-wide forward model (7,345 fibers) and two local fits, with the petal term from that model or a free constant per tile--petal, give E-ridge peaks of 1.69--1.83\,R at P1 and 1.34--1.68\,R at P3 (Figure~\ref{fig:desi_fibers}i), a range that includes the southern-ridge median of 1.62\,R (Section~\ref{sec:brightness}). The spread between models (0.14 and 0.34\,R) exceeds the statistical errors.

DR1 lists no guide-camera acceptance for the two survey-validation tiles; we adopted program medians, $f=0.419$ (bright) and 0.608 (dark). The same ridge then reads $0.94\pm0.06$\,R in tile 80644 against $1.70\pm0.07$\,R in tile 80683 at P1, and $0.99\pm0.07$ against $1.53\pm0.07$\,R at P3, whereas the calibrated fluxes have ratios of $0.80\pm0.06$ and $0.94\pm0.08$ (1.45 expected for the adopted acceptances). Tiles with measured acceptance give $1.62\pm0.16$\,R near P1 (43 fibers). The brightness near P1 and P3 is therefore uncertain by $\approx$20\%, more than the $\pm8\%$ of Section~\ref{sec:data}; the adopted medians enter 46\% of the fibers at Dec $<19.5^\circ$. W26 Table~3 remains 6.2 and 6.6 times the model sky brightness.

\subsection{Line ratios}

We use three estimators: the regression of Section~\ref{sec:data} restricted to each region, with an intercept or with tile--petal means removed, and Gaussian fits to the ON$-$OFF stacks in energy units. Uncertainties are bootstraps over fibers, with floors from the 17 line-free windows (ratio rms 0.029--0.040 at P1 and 0.018--0.022 at P3) and from 300 random OFF$-$OFF splits.

Both regions show an \Oiii{} doublet ratio consistent with 2.98 and \Oii{} at 7--11$\sigma$ (Figure~\ref{fig:desi_fibers}j,l). \Oii/\Oiii{} is within $1\sigma$ of the $1.11$--$1.16$ ($\pm0.08$) measured at Dec $<19.5^\circ$ (Table~\ref{tab:ratios}) at P1 and 1--2$\sigma$ lower at P3. \Hb{} and \ion{He}{2} are undetected; the limits are weaker than the filament-wide ones, except \Oiii/\Hb{} $>18$ at P3.

\Ha{} is positive (0.12--0.19\,R, 5008\,\AA{} equivalent) in both ON and OFF stacks, so the red-line entries limit only emission that follows the \Oiii{} ridge (Figure~\ref{fig:desi_fibers}k). \Nii~$\lambda$6583 at P3 is 0.045--0.068 of \Oiii{} (2.0--3.4$\sigma$; 2.8--4.7$\sigma$ against red null windows alone); the data cannot separate filament \Nii{} from WIM contrast between ridge and off-ridge fibers.

\subsection{Limitations}

The petal sky model was not inspected; its mean is modeled from fiber positions, and near P3 this term ($\approx$0.97\,R) is as large as the signal. Two tiles and at most six tile--petals dominate each region, so the bootstraps over fibers (the regions are too small for the block bootstrap of Section~\ref{sec:data}) do not sample errors shared by a petal.

\bibliography{references}{}
\bibliographystyle{aasjournal}

\end{document}